\documentclass{sig-alternate-10pt}

\usepackage{color}
\usepackage{booktabs} 
\usepackage{color}
\usepackage{subfigure}
\usepackage{url}
\usepackage{listings}

\usepackage{authblk}
\usepackage{color}
\newcommand{\para}[1]{{\vspace{4pt} \bf \noindent #1 \hspace{6pt}}}
\newenvironment{packed_itemize}{
\begin{list}{\labelitemi}{\leftmargin=1.em}
  \setlength{\itemsep}{3pt}
  \setlength{\parskip}{0pt}
  \setlength{\parsep}{0pt}
  \setlength{\headsep}{0pt}
  \setlength{\topskip}{0pt}
  \setlength{\topmargin}{0pt}
  \setlength{\topsep}{0pt}
  \setlength{\partopsep}{0pt}
}{\end{list}}

\definecolor{change}{rgb}{0,0,0}
\definecolor{codegreen}{rgb}{0,0.6,0}
\definecolor{codegray}{rgb}{0.5,0.5,0.5}
\definecolor{codepurple}{rgb}{0.58,0,0.82}
\definecolor{backcolour}{rgb}{1,1,1}
\definecolor{codeblack}{rgb}{0,0,0}

\lstdefinestyle{mystyle}{
	backgroundcolor=\color{backcolour},   
	commentstyle=\color{codeblack},
	keywordstyle=\color{codeblack},
	numberstyle=\tiny\color{codeblack},
	stringstyle=\color{codeblack},
	basicstyle=\footnotesize,
	breakatwhitespace=false,         
	breaklines=true,                 
	captionpos=b,          
	frame=single,          
	keepspaces=true,                 
	numbers=left,                    
	numbersep=5pt,                  
	showspaces=false,                
	showstringspaces=false,
	showtabs=false,                  
	tabsize=1,
	float=t,
	keywordstyle=\color{blue}\bfseries,
	floatplacement=tbp,
	belowcaptionskip=-0.5in,
	columns=flexible,
}

\begin{document}
\title{Trimming Mobile Applications for Bandwidth-Challenged Networks in Developing Regions}
\author[*]{Qinge Xie}
\author[*]{Qingyuan Gong}
\author[*]{Xinlei He}
\author[*]{Yang Chen}
\author[*]{Xin Wang}
\author[$\dag$]{Haitao Zheng}
\author[$\dag$]{Ben Y. Zhao}
\affil[*]{School of Computer Science, Fudan University, \authorcr \{qgxie17, gongqingyuan, xlhe17, chenyang, xinw\}@fudan.edu.cn}
\affil[$\dag$]{Department of Computer Science, University of Chicago, \{htzheng, ravenben\}@cs.uchicago.edu}

\maketitle

\begin{abstract}
Despite continuous efforts to build and update network infrastructure, mobile devices in developing regions continue to be constrained by limited bandwidth. Unfortunately, this coincides with a period of unprecedented growth in the size of mobile applications. Thus it is becoming prohibitively expensive for users in developing regions to download and update mobile apps critical to their economic and educational development. Unchecked, these trends can further contribute to a large and growing global digital divide.

Our goal is to better understand the source of this rapid growth in mobile app code size, whether it is reflective of new functionality, and identify steps that can be taken to make existing mobile apps more friendly bandwidth constrained mobile networks. We hypothesize that much of this growth in mobile apps is due to poor resource/code management, and do not reflect proportional increases in functionality. Our hypothesis is partially validated by mini-programs, apps with extremely small footprints gaining popularity in Chinese mobile networks. Here, we use functionally equivalent pairs of mini-programs and Android apps to identify potential sources of ``bloat,'' inefficient uses of code or resources that contribute to large package sizes. We analyze a large sample of popular Android apps and quantify instances of code and resource bloat. We develop techniques for automated code and resource trimming, and successfully validate them on a large set of Android apps. We hope our results will lead to continued efforts to streamline mobile apps, making them easier to access and maintain for users in developing regions.

\end{abstract}

%
%



\section{Introduction}

For most users in developing regions today, bandwidth for mobile devices is
still a very limited resource. Most users rely on cellular networks dominated
by older infrastructure (2G or 2G+EDGE)~\cite{Mariya_mobisys13}, often
augmented by satellite or low-bandwidth, long-distance wireless links. The
result is overall poor quality of Internet access~\cite{Zambia_it15}, with
bandwidth of only hundreds of kbps~\cite{internet_usage_www11,matthee_ictd07}. Despite efforts ranging from
long-distance wireless links~\cite{wildnet,wildnet2,ldwifi}, localized
cellular networks~\cite{Mariya_mobisys13} to affordable commodity WiFi
hotspots~\cite{keith}, growth in mobile bandwidth is still slow.
Actual bandwidth available to users is often constrained by multiple factors
including cost, last mile congestion, and limited access to backhaul links.

Unfortunately for users in developing regions, mobile applications worldwide
are growing in size at an unprecedented pace, in part due to the growth of
cheap or unlimited cellular plans. For example, bandwidth required to
download the top 10 most installed U.S. iPhone apps ({\em e.g.}, Facebook,
Uber, YouTube) has grown by an order of magnitude from 164MB in 2013 to about
1.9GB in 2017~\cite{average_ios_size_2017}. In the US, these ``growing'' app
sizes mean that software updates now account for a big chunk of cellular
bandwidth across the country~\cite{appupdate,appupdate2}.  Unsurprisingly,
studies already show that larger mobile applications lead to stability or
usability problems on constrained
networks~\cite{Ihm_nsdr10,Isaacman_www11,Media_Connected_dev15}.

In concrete terms, this means that users in developing regions will find it
difficult or impossible to access some of the most popular mobile apps
critical to economic and educational development, despite studies that show
tremendous impact from mobile apps on agriculture, health and
education~\cite{agriculture_dev13,health_app_family_dev16,education_dev15}.
For example, Duolingo, the popular app for learning foreign languages, has an install package 
of size 20MB, and as of May 2018, provides frequent updates with bug fixes
that require a full download of the app each week.  Khan Academy, the popular
online education app, has an install package of 22MB, and updates its software roughly once
every 2 weeks. Other popular applications also have surprisingly large install packages. CodeSpark Academy is at 59MB, 
Facebook is at 65MB, and Uber takes 61MBs to download. Even simple apps from American Airlines 
and McDonalds require 83MB and 43MB to download respectively. 

At first glance, these trends seem to predict a widening digital divide where developing regions are losing access to critical parts of the mobile app ecosystem.  
But is the situation truly as dire as it seems? Intuitively, it seems
unlikely that this staggering growth in size of mobile apps is truly driven
by growth in functionality. What factors other than functionality are
contributing to this growth? Perhaps more importantly, how much of this growth is
truly necessary for mobile apps, and how much can be traded off in return for
app sizes more friendly to bandwidth-constrained networks?

In this paper, we describe our efforts to answer these questions, through
a deeper understanding of factors that contribute to the accelerating growth
in the size of mobile applications. We use a variety of empirical tools and
techniques to break down mobile applications\footnote{Given the dominance of
  Android smartphones in developing regions~\cite{androidvsios}, we focus
  exclusively on Android apps in this study.}, and find that for a large
number of mobile apps across all categories, much of the increases in app
size can be attributed to the casual inclusion of both resource files and
linked software libraries, much of which is never called by the mobile app
code. These findings suggest it is possible to produce significantly smaller
mobile apps suitable for bandwidth-limited networks by trimming unreferenced
library code and making bandwidth-aware tradeoffs with resource files.

Our hypothesis is partially validated by the popularity of mini-programs~\cite{miniprogs,mini_program} or mini-apps~\cite{miniapps},
apps with extremely small footprints that run
on top of mobile platforms in China. While some of them have reduced functionality compared to their mobile app counterparts, others retain similar functionality but at a small fraction of the package footprint to meet the resource constraints imposed by their parent apps, {\em e.g.}, WeChat and Alipay\footnote{WeChat is the dominant mobile messaging and social platform in China (1B+ users), and AliPay is  the dominant mobile payment system in China.}. For example, WeChat limits its mini-programs to an installation package of 2MB. Tight limits on app package size allow these platforms to adopt a load-on-demand approach to app discovery, where users can discover and run mini-programs ``instantly'' with negligible delay.


In our study, we analyze a large selection of Android apps to understand the different software components and their contributions to overall app size. In the process, we identify multiple types of ``code bloat'' that can be trimmed from app packages without impact to functionality, including unreferenced third-party code and redundant resource files. We also develop generalizable techniques for trimming code bloat from existing mobile apps. Our results show that combined with compacting images and documentation, eliminating code bloat can significantly reduce the package size of many existing apps, making them much more usable in bandwidth-constrained environments such as developing regions.

We summarize our key contributions as follows:
\begin{itemize}
\item We perform code analysis on a selection of popular Android mobile apps in detail to understand potential sources of code bloat. To identify potential benefits of mini-program platforms like WeChat, we implement a mini-program from scratch with identical functionality as an existing Android app, and analyze them to understand sources of app size discrepancy. We identify the size of linked libraries as a key contributor.
\item We perform detailed analysis of 3200 of the highest ranked Android apps on Google Play app store, and confirm that linked libraries are a dominant factor in their overall app size. We use static analysis to identify unreferenced methods and classes, and use automated tools to remove unreferenced code from these apps. We run a state exploration tool to estimate usage of resources, and find that significant pruning is possible for both code and resources.
\item Based on our results, We introduce the idea of a streamlined mobile app platform for bandwidth-constrained networks, and propose an automated process to reduce code bloat in existing Android apps.  It packs commonly used APIs into a single library, allowing for reuse and minimizing per-app package size. This platform could be deployed by mobile phone OS developers like Google Android or by app (platform) developers like Tencent.
\end{itemize}

We note that Google is already spearheading developer initiatives for more lightweight mobile phone design in their ``Building for Billions'' initiative\footnote{\url{https://developer.android.com/topic/billions/}}. Our effort is complementary, in that we focus more on the question of bloat for existing Android Apps, and how they could be retrofitted to perform better in constrained networks. We hope this work leads to continued efforts on code trimming for existing mobile apps, and provides additional support for lightweight development efforts like Building for Billions.
\section{Background and Related Work}
\label{sec:miniapps}

\para{Mini-programs and Lightweight Apps.} We begin by providing some background on the development of mini-programs by WeChat and Alipay. Mini-programs provide an extreme example of what is possible if code size were prioritized over all other concerns.

WeChat and Alipay are the two largest Internet apps in China, in both users and influence. WeChat is a ubiquitous messaging platform with a person-to-person mobile payment component that has become more accepted in China than cash. Alipay is a Chinese financial conglomerate that dominates the mobile payment market.
While WeChat might be analogous to a union of Facebook and Venmo, Alipay might be a combination of PayPal and Amazon.

WeChat's goal for its mini-programs is to introduce users to new apps in real time, often scanning a QR code to instantaneously install and run a mini-program~\cite{miniprogs}. Thus mini-programs have to be extremely small in size. The current limit is 2MB for the entire installation package, including code, libraries and resource files. In reality, this development effort elevates WeChat to an app ecosystem capable of competing against Apple's AppStore or Google Play (which is blocked in China), and WeChat encourages its users to bypass traditional app stores entirely. Since launching in January 2017, WeChat now runs 580,000 mini-programs, compared to 500,000 mobile apps published by Apple's AppStore between 2008 and 2012~\cite{minibattle}. 

The popularity of WeChat mini-programs led to a competing effort from AliPay, which launched their own ``mini-app'' platform in October 2017~\cite{miniapps}. AliPay's platform also set 2MB as the limit for install package size, with an internal limit of 10MB for storage. Given their similarity, we focus our analysis on WeChat mini-programs, and use mini-programs in the paper tp refer to WeChat mini-programs.

Mini-programs and mini-apps also share properties with several other alternative lightweight app platforms. {\em Web Applications} are accessible via the mobile device's Web browser like Safari or Chrome. The app runs on the server and visual elements are sent to the mobile device. Google announced {\em Android Instant Apps}~\cite{instant_app} at Google I/O 2016. Users can tap to try an instant app without installing it first. An instant app only loads portions of the app on an on-demand basis. We note that our goal is to streamline mobile apps that run on the device, which differs from these platforms, since they both rely on network infrastructure.

\para{Redundant Code And Library Usage.}
Code redundancy is a common phenomenon in software engineering. It is often the case that the client code uses an API in an inefficient way~\cite{api_ase09}. A variety of solutions have been proposed to help developers detect redundant code in their projects during the development process. {\color{change} Derr \emph{et.al}~\cite{library_ccs18} showed that extensive use of third-party libraries can introduce critical bugs and security vulnerabilities that put users' privacy and sensitive data at risk.}
Kawrykow \emph{et.al}~\cite{api_ase09} developed a tool to automatically detect redundant code and improve API usage in Java projects.  Lammel \emph{et.al}~\cite{api_sac11} designed an approach to deal with large-scale API-usage analysis of open-source Java projects. This technique helps with designing and defending mapping rules for API migration in terms of relevance and applicability. 
Wang \emph{et.al}~\cite{api_msr13} proposed two qualified metrics for mining API-usage patterns, i.e., succinctness and coverage. They further proposed a novel approach called Usage Pattern Miner (UP-Miner), for mining succinct and high-coverage usage patterns of API methods from source code.

\section{Mini-Programs vs. Mobile Apps}
\label{section3}

The proliferation of mini-programs in China proves that for hundreds of thousands of mobile apps, their core functionality could be implemented in an extremely compact form. The question is, what are the tradeoffs necessary to obtain that compact implementation? What accounts for the difference in code size; is it more efficient code, or were there significant losses in functionality?
In this section, we search for an answer by comparing mini-programs to their Android app counterparts, in both app content/features and code structure. We describe two detailed illustrative examples, and then present results of empirical analysis of 200 of the most popular mini-programs and their Android app counterparts.



\begin{table}[t]%
	\caption{\small Constraints on Mini-Programs}
	\centering
	\resizebox{1\columnwidth}{!}{
	\begin{tabular}{ | l |l|}
		\toprule[2pt]
		\textbf{Restricted Item}           & \textbf{Description} \\
		\midrule[1pt]
		\midrule
		\textbf{Page Depth}              & $\leq$ 5\\
		\textbf{Package Size}  & $\leq$ 2MB  (8MB if using subpackage)\\
		\textbf{Local Storage}              & $\leq$ 10MB per user (WeChat)\\
		\textbf{API Usage}                  & WeChat offers certain APIs for development \\
		\bottomrule
	\end{tabular}
	}
	\label{tab:one}
\end{table}%

\subsection{Overview}
Mobile apps face little design constraints other than size limit. Google Play~\cite{google_play_limit} limits Android apps by 100MB (Android 2.3+) but allows two expansion files totalling up to 4GB; iOS allows 4GB for any submission to its App store~\cite{apple_store_limit}. A recent measurement study in 2017 shows that the average sizes of iOS and Android apps are 38MB and 15MB, respectively~\cite{average_ios_size_2017}. 


Mini-programs must abide by a number of restrictions, in their page depth (max number of links necessary to reach any page), local storage, installation package size, and API usage, which we summarize in Table~\ref{tab:one}\footnote{We omit Game mini-programs in our table, which are granted 50MB for local storage.}. WeChat 6.6.0+ allows the usage of subpackages in mini-programs, but still limits the size of a single package to 2MB and the total sum of all packages to 8MB. 
To meet these constraints, developers often simplify app features and user interface (UI) elements in mini-program versions of their mobile app.  Furthermore, mini-programs use JavaScript rather than Java (used by Android apps). We note that while code size across these two languages can vary (up to 20\%~\cite{javascriptcodesize}), any syntactical differences are unlikely to be meaningful. This is because APKs and mini-programs are stored under compression, and compression algorithms are likely to be more efficient on more verbose representations. 


Using a common unpack tool~\cite{wechat_unpack} to parse mini-programs, we are able to compare the overall code composition of mini-programs and their Android counterparts (see Table~\ref{tab:three}). Implemented using JavaScript, a mini-program's main program code resides in \emph{app-service.js}, which is analogous to the Java Bytecode file in the Android APK ({\em i.e.\/}, Android Dex file). There is a common setting file (\emph{app-con.json}), analogous to \emph{AndroidManifest.xml} file in the Android APK. Each page appeared in the mini-program is registered in the common setting file, and there is a folder for every page to include its CSS configuration. The main page design code (HTML) is packaged in \emph{page-frame.html}, analogous to Layout files in Android.




\begin{table}[t]
	\caption{\small Common structure of mini-program installation package.}
	\vspace{-0.2in}
	\begin{center}
		\resizebox{1\columnwidth}{!}{
		\begin{tabular}{|l| p{0.4\columnwidth} | p{0.4\columnwidth} |}
			\toprule[2pt]
			\textbf{File / Folder Name} & \textbf{Description} & \textbf{Correspondence in Android APK} \\
			\midrule[1pt]
			\midrule
			app-service.js & Main program logic codes & Android Dex files\\
			app-config.json & Common settings file & AndroidManifest.xml \\
			page-frame.html	& The integration of layout files of all pages & Layout files\\
			Pages folder & CSS configuration files of every page & Layout files \\ 
			Other folders or files & Other resource files & Other folders or files \\
			\bottomrule
		\end{tabular}
		}
		\label{tab:three}
	\end{center}
\end{table}%

\subsection{Code Package Composition}
Next, we perform detailed comparisons between two mini-programs and their Android counterparts. One program is a popular Android app whose developers implemented their own mini-program, and the other is an Android app for which we implemented a mini-program that precisely replicated its functionality. 

\begin{figure}[t]
		\centering
	\subfigure[Mini Program]{
		\label{fig:one:a} 
		\includegraphics[width=0.15\textwidth]{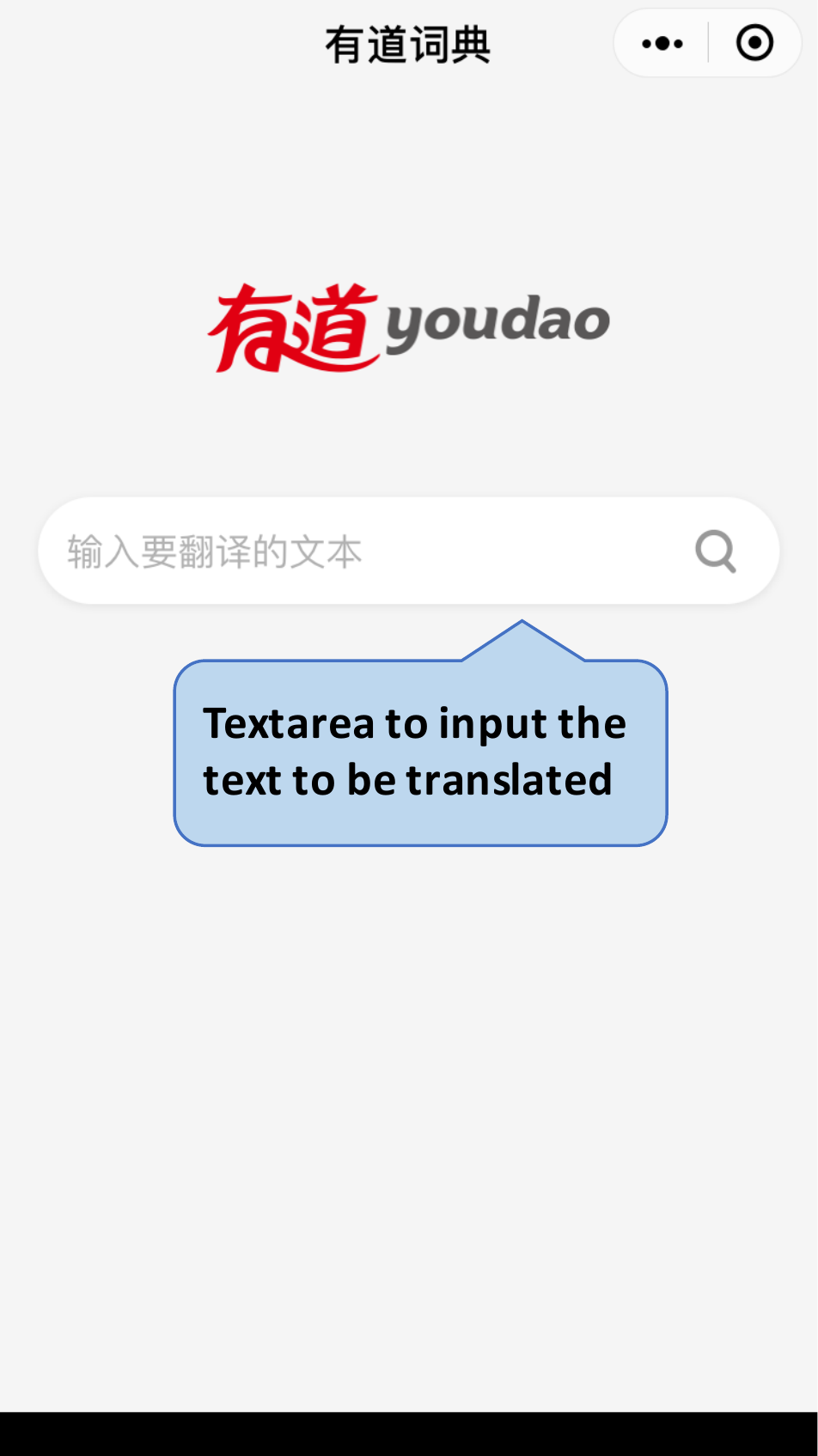}}
	\hspace{0.2in}
	\subfigure[Android App]{
		\label{fig:one:b} 
		\includegraphics[width=0.15\textwidth]{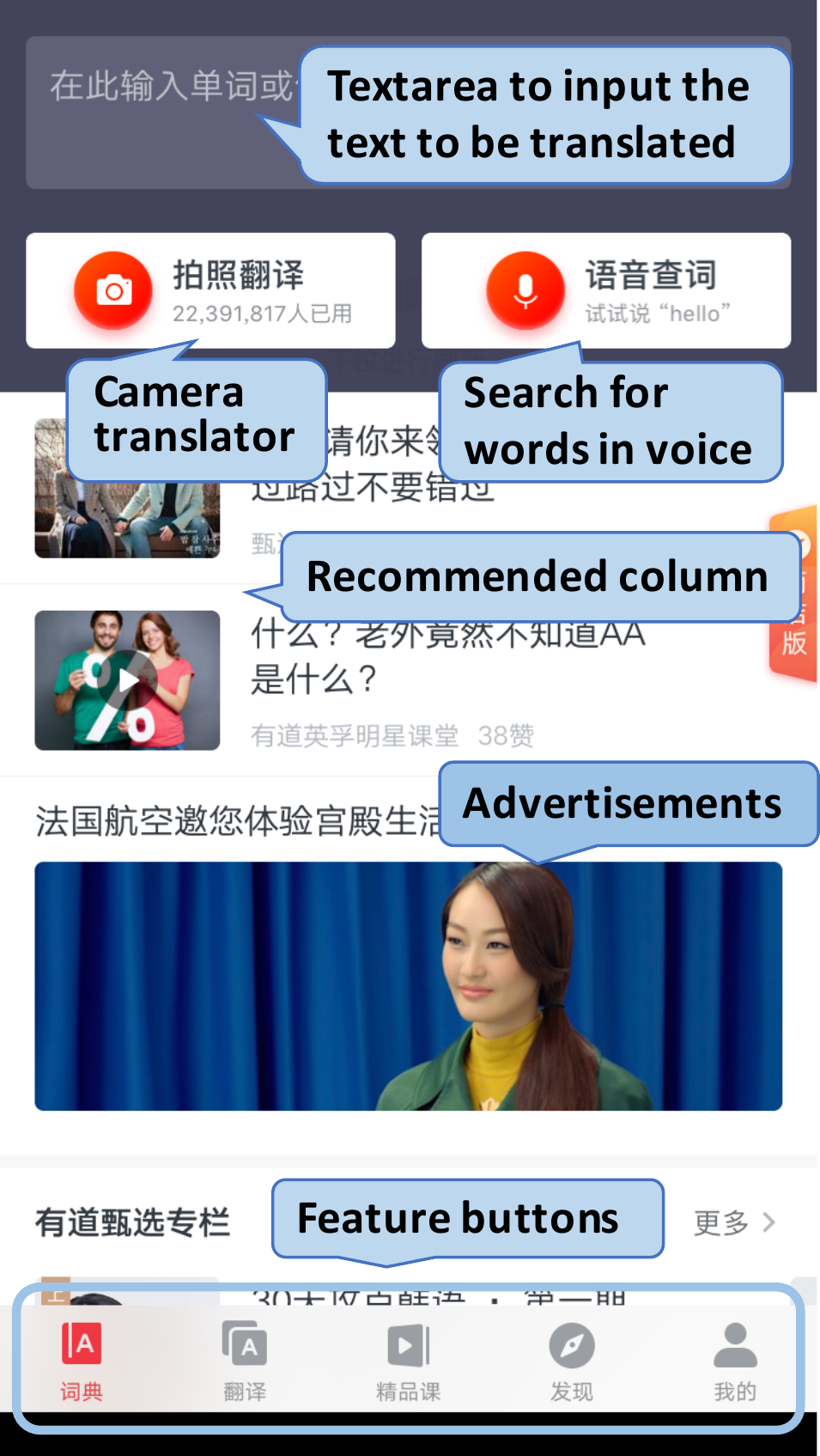}}
		\vspace{-0.05in}
	\caption{\small NetEase YouDao Dictionary App}
	\label{fig:one} 
	\end{figure}

\para{Example 1: YouDao Translation Dictionary.} {\em YouDao} is a very popular multi-language (7+ languages) translation dictionary app, which has an official mini-program version. As shown in Figure~\ref{fig:one}, the mini-program includes the app logo and an input box for translation, while the full app provides a more sophisticated UI and several extra features (camera translator, human translator, audio word search, courses and vocabulary book). 

	\begin{figure}[t]
		\centering
		\subfigure[mini-program]{
			\label{fig:two:b} 
			\includegraphics[width=0.2\textwidth]{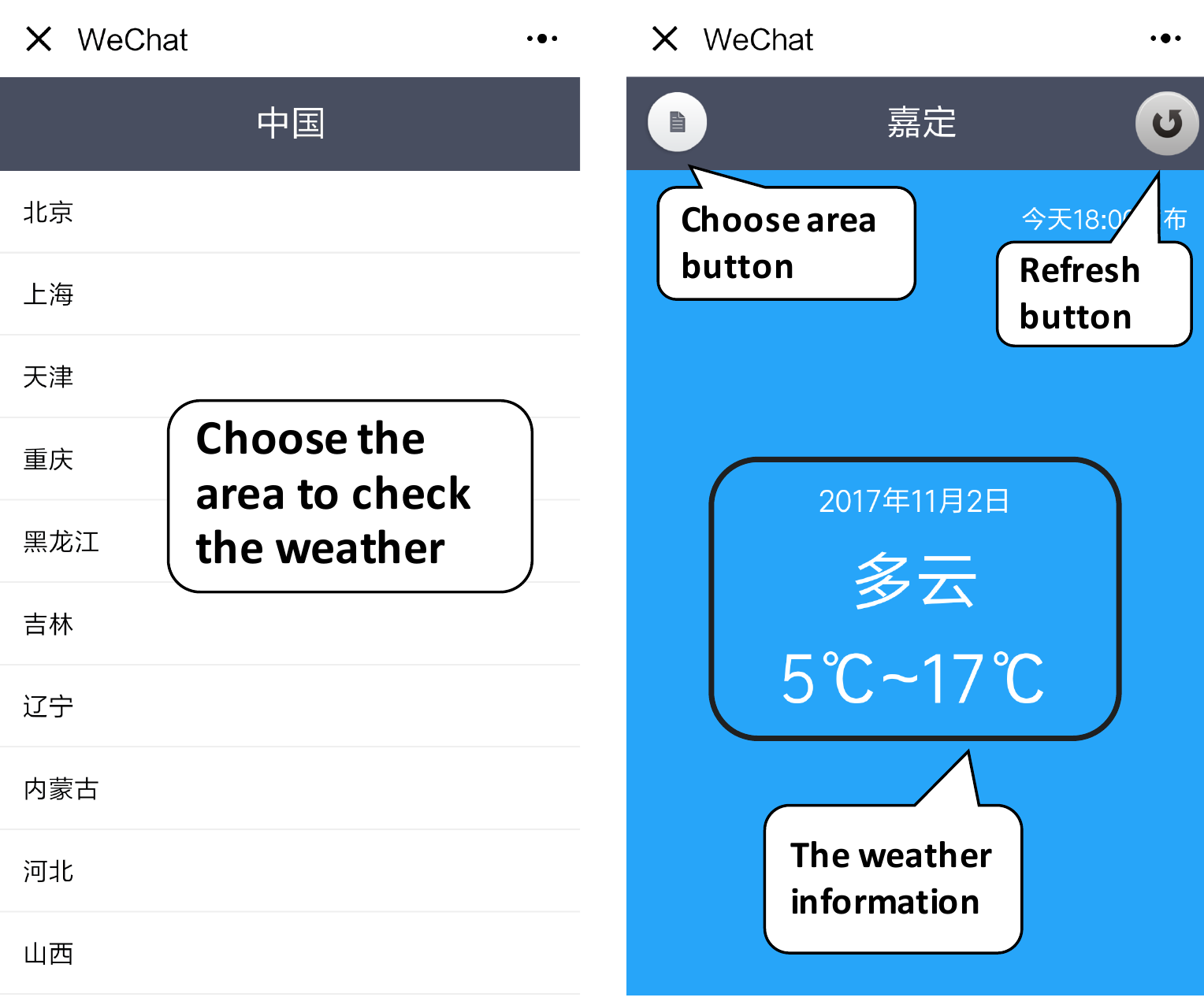}}
		\hspace{-0.05in}
		\subfigure[Android App]{
			\label{fig:two:a} 
			\includegraphics[width=0.19\textwidth]{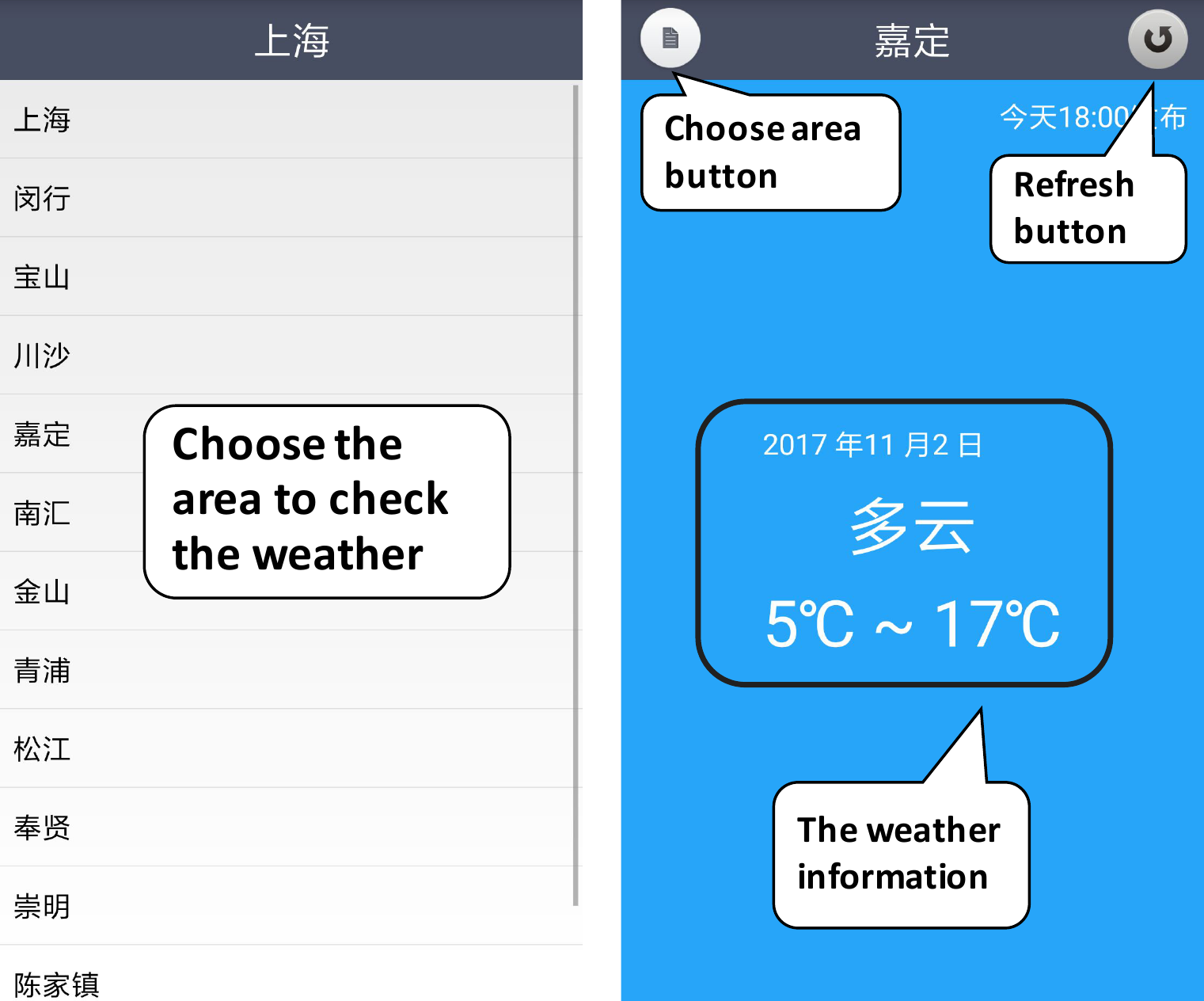}}
				\vspace{-0.15in}
		\caption{\small Today Weather App}
		\label{fig:two} 
\end{figure}
	
\begin{table*}[t]
	\caption{\small Package analysis of mini-program and Android app version of YouDao Dictionary.}	\label{tab:four}
	\centering
		\resizebox{1.4\columnwidth}{!}{
	\begin{tabular}{|p{0.4\columnwidth} |l r l| l r l|}
	\toprule[2pt]
	& \textbf{Android App}           &                         &          & \textbf{Mini Program}& &\\
	\midrule[1pt]
	\midrule[1pt]
	\textbf{Total installation package size}      & \emph{APK size}          & 47.2MB  &   &     \emph{WXAPKG size}      & 0.215MB&\\
				             & \emph{Folder size}       & 60.6MB &    & \emph{Folder size}     & 0.232MB&\\
	\midrule
	\textbf{Res Resources}   & \emph{Images} & 12.25MB  &(21.21\%) & \emph{Images} & 0.161MB &(69.40\%)  \\
				 		 & \emph{Layout files}  &2.47MB &(4.08\%)   & \emph{HTML/CSS} & 0.049MB & (21.12\%)  \\
				 		  \midrule
				\textbf{Assets	Resources} & \emph{Assets}  & 17.43MB& (28.76\%) & - & -&  \\
						 				 \midrule
				\textbf{C++ Library} & \emph{Lib} & 15.3MB &(25.25\%) & - & -& \\
								 \midrule
				\textbf{Procedure Code} & \emph{Android Dex file} & 10.5MB &  (17.33\%) & \emph{app-service.js} & 0.015MB&(6.47\%) \\

				\bottomrule[1pt]
			\end{tabular}
			}
\end{table*}%

\begin{table*}[t]
	\caption{\small Package analysis of mini-program and Android app version of Today Weather.}
	\centering
		\resizebox{1.4\columnwidth}{!}{
		\begin{tabular}{|p{0.4\columnwidth} |l r l|l r l|}
			\toprule[2pt]	
			& \textbf{Android App} & & &  & \textbf{Mini Program} &\\
			\midrule[1pt]
			\textbf{Total installation package size}& \emph{APK size} & 527 KB & & \emph{WXAPKG size}& 82.1 KB &\\
			& \emph{Folder size} & 1452.01 KB  &   & \emph{Folder size} & 81.84 KB&\\
			\midrule
			\textbf{Res Resource} & \emph{Images}  & 65.63 KB & (4.52\%) & \emph{Images} & 30.17 KB &(36.88\%)  \\
			& {\em Layout file}  & 6.80 KB& (0.47\%) & \emph{HTML/CSS}& 42.82 KB & (52.32\%)  \\
			\midrule
			\textbf{Assets Resource} & \emph{Assets}  & 0 KB& (0\%) & - & - & \\
			\midrule
			\textbf{C++ Library} & \emph{Lib} & 0 KB &  (0\%) & - & -& \\
			\midrule
			\textbf{Procedure Code}& \emph{Android Dex file} & 1361.92 KB& (93.80\%)  & \emph{app-service.js} & 8.38 KB& (10.23\%) \\
			\midrule
			\textbf{Configuration File} & \emph{AndroidManifest.xml} & 2.42 KB &(0.17\%)  & \emph{app-config.json} & 0.47 KB& (0.57\%) \\
			\bottomrule
		\end{tabular}
		}
		\label{tab:five}
\end{table*}%




Table~\ref{tab:four} lists the per-component code comparison between the mini-program and the Android app (we discuss these components in more detail later in \S\ref{sec:library}). We see that the total footprint for the mini-program is 0.2MB, compared to 47.2MB for the Android app (219X smaller!). Across each analogous component, the mini-program version is smaller by at least a factor of 100!
This is an example where all aspects of the Android app were compacted to generate its matching mini-program. While the core functionality remains, some non-core features were cut and the UI was simplified. 

\para{Example 2: Today Weather App.} In an app like YouDao Translation Dictionary, the developer made specific tradeoffs in choosing which areas to trim. We wanted to find a more controlled example where full functionality was preserved in the mini-program, so we could better understand the impact of compressing components unrelated to core features. The only way to ensure a true apple-to-apple comparison was to implement a mini-program ourselves, and ensuring that functionality of the Android app was preserved perfectly.

We found a reasonably sized Android app with no matching mini-program, the Today Weather app, an Android app from the 360 app store, which provides city-wise weather conditions in China. Figure~\ref{fig:two} shows the {\em Today Weather} app and its mini-program version.

To build a matching mini-program, we first decompiled the Android app from its Dalvik bytecode into Java bytecode using the well-known tool dex2jar~\cite{dex2jar}.  Since the app's program code, logic and function calls are all accessible, we replicated them completely with one minor exception\footnote{\small Due to security restrictions on mini-programs (HTTPS requests only), we had to change HTTP requests in the Android app to HTTPS requests in the mini-program. This minor change should have zero impact on the outcome of our analysis.}. We also made sure that resource files like images were also identical to their Android counterparts. We tested our mini-program thoroughly to confirm that it offers the same interfaces, program logic, resources, and function calls. 

Table~\ref{tab:five} lists the package analysis of the two programs. While providing the same feature, program logic, resources and network requests, the mini-program still achieves significant reduction in app size: 82.1KB vs. 527KB (compressed) or 1452KB (uncompressed) for the Android app, mapping to a factor of 6x to 18x. 

A closer look shows that while the procedure code file in the Android app occupies 1.36MB (93.80\% of the package after decompressing the APK), the corresponding code file in the mini-program is only 8.38KB, which is 160 times smaller! In fact, the procedure code file of the Android app is dominated by the Java library file, which takes up 
95.59\% of the code space. At least in this example, we find that we can significantly trim an Android app while preserving functionality and content (images and features). The key here is streamlining the procedure code, and more specifically, its Java library file. 



\para{Summary of Findings.} These two examples show that developers can achieve drastic reductions by shrinking all components of mobile apps, including features and content (e.g. images). But even while preserving features and content, we can achieve significant savings by handling libraries more efficiently. We revisit this in more detail in Section~\ref{sec:library}.

\subsection{Pair-wise Package Analysis}

Now that we have a high level sense of potential areas for trimming mobile apps, we extend our comparison of Android apps and mini-programs to 200 of the most popular app pairs. The goal is to get a high level picture of how code, resources (e.g. images) and functionality compare across popular mobile apps, and how much room for trimming each category represents.



\para{Dataset.} We build a list of popular mini-programs from the monthly list of top-100 mini-programs published by aldwx.com\footnote{\url{https://www.aldwx.com/} is a third-party statistics platform for WeChat mini-programs.}. We include all mini-programs ever to appear on the top-100 list before October 2018.
For each mini-programs, we identify their corresponding Android app counterpart using a combination of application name, developer, official identification, and manual confirmation. Our final dataset includes the 200 popular mini-programs and their official Android app counterparts.

\begin{figure*}[t]
	\centering
	\setlength{\abovecaptionskip}{0.06in} 
	\setlength{\belowcaptionskip}{-0.2in} 
	\subfigure[Installation package size]{
		\label{fig:ex1:a} 
		\includegraphics[width=2.3in]{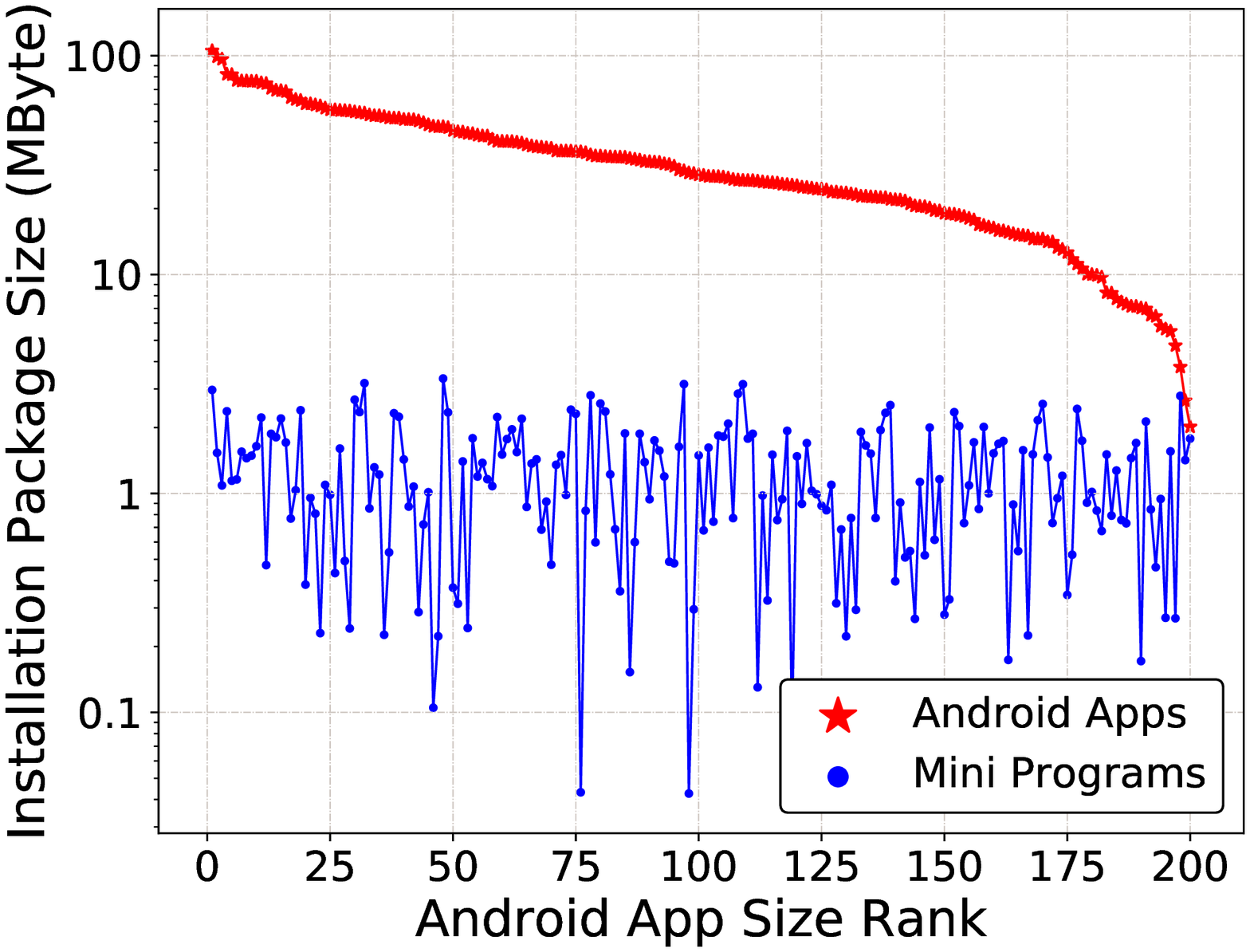}}
	\hspace{-0.2in}
	\subfigure[Image size and image count]{
		\label{fig:ex1:b} 
		\includegraphics[width=2.3in]{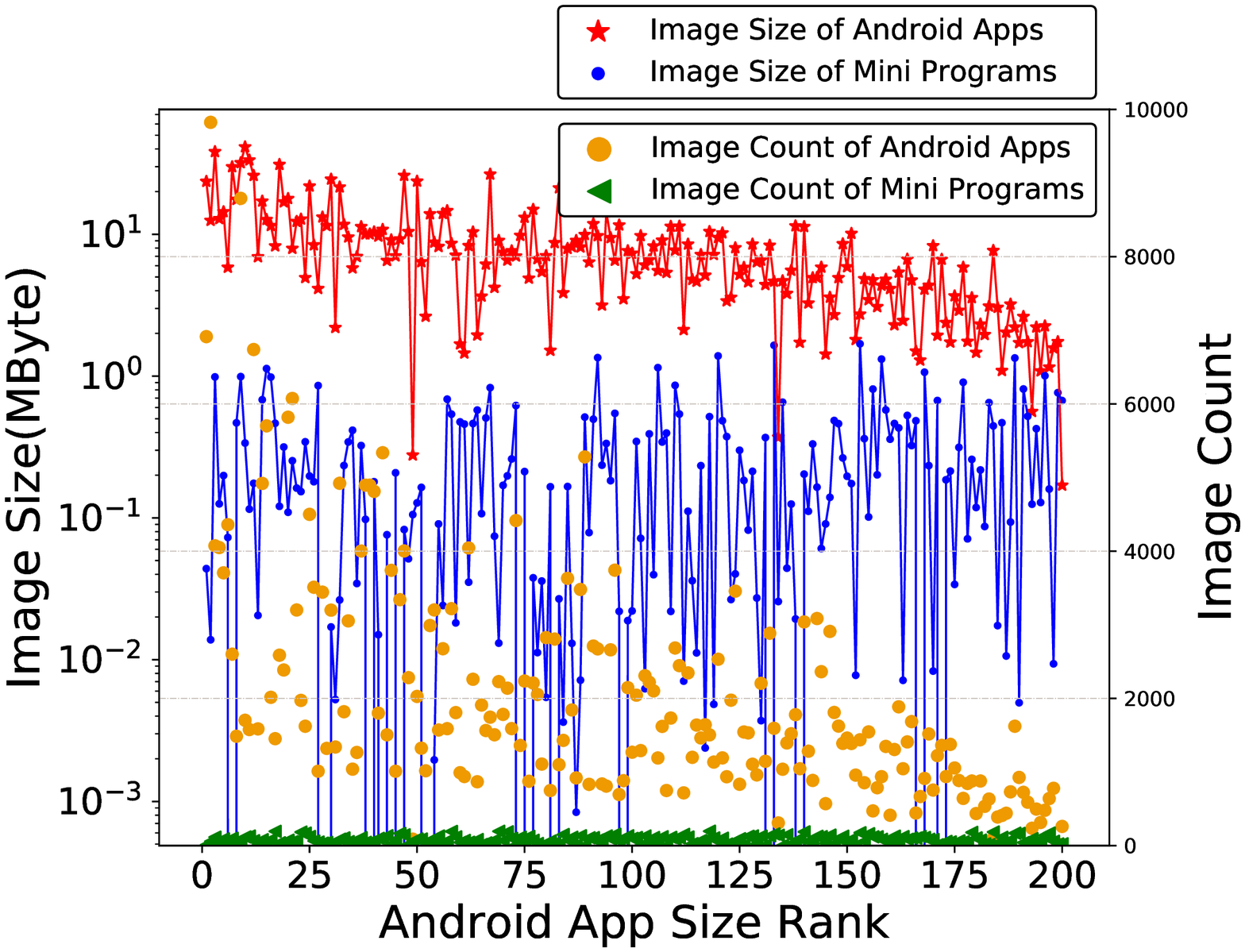}}
	\subfigure[Features (count of pages/activities)]{
		\label{fig:ex1:c} 
		\includegraphics[width=2.3in]{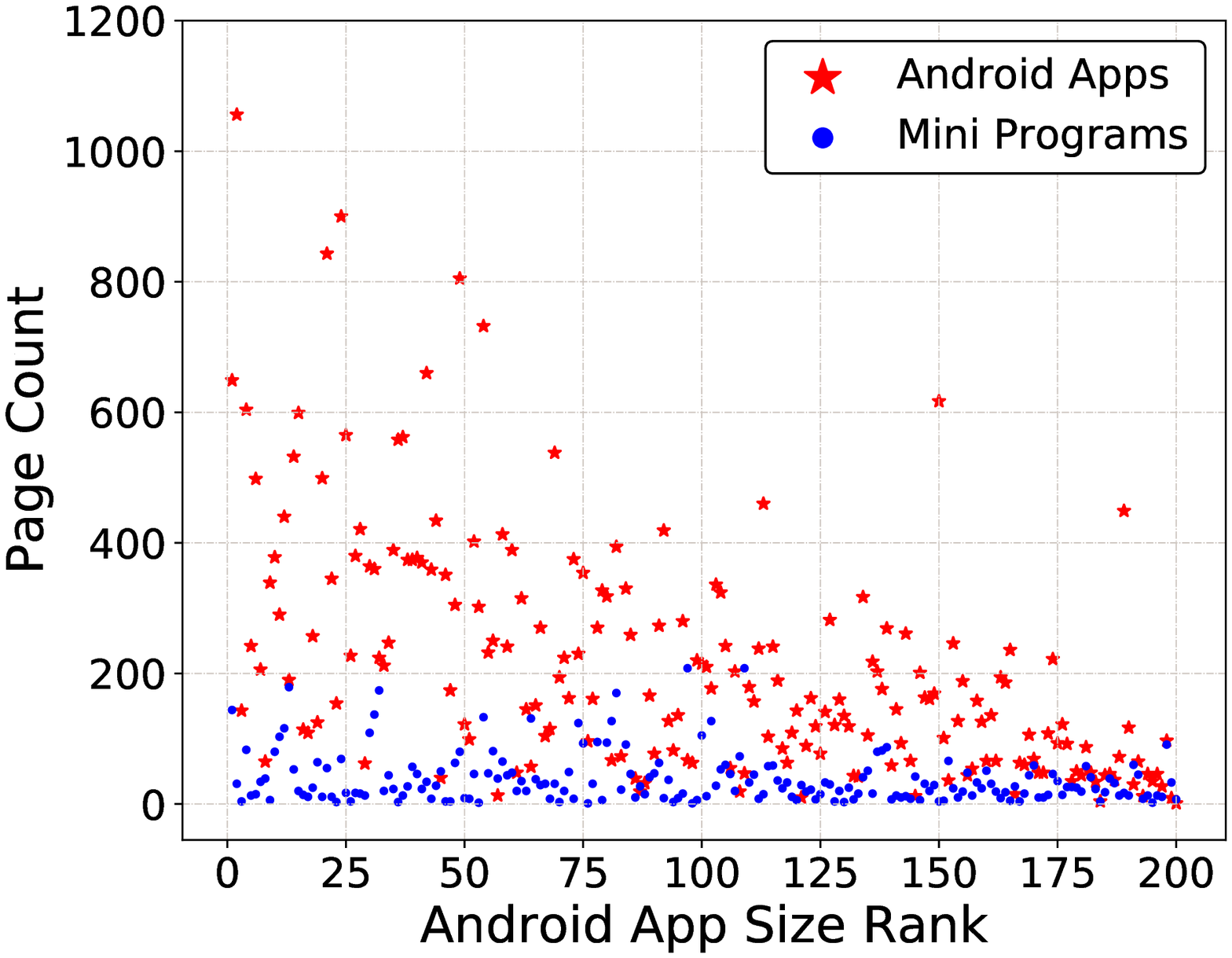}}
	\subfigure[Package composition: Android]{
 		\label{fig:ex1:e} 
 		\includegraphics[width=2.8in]{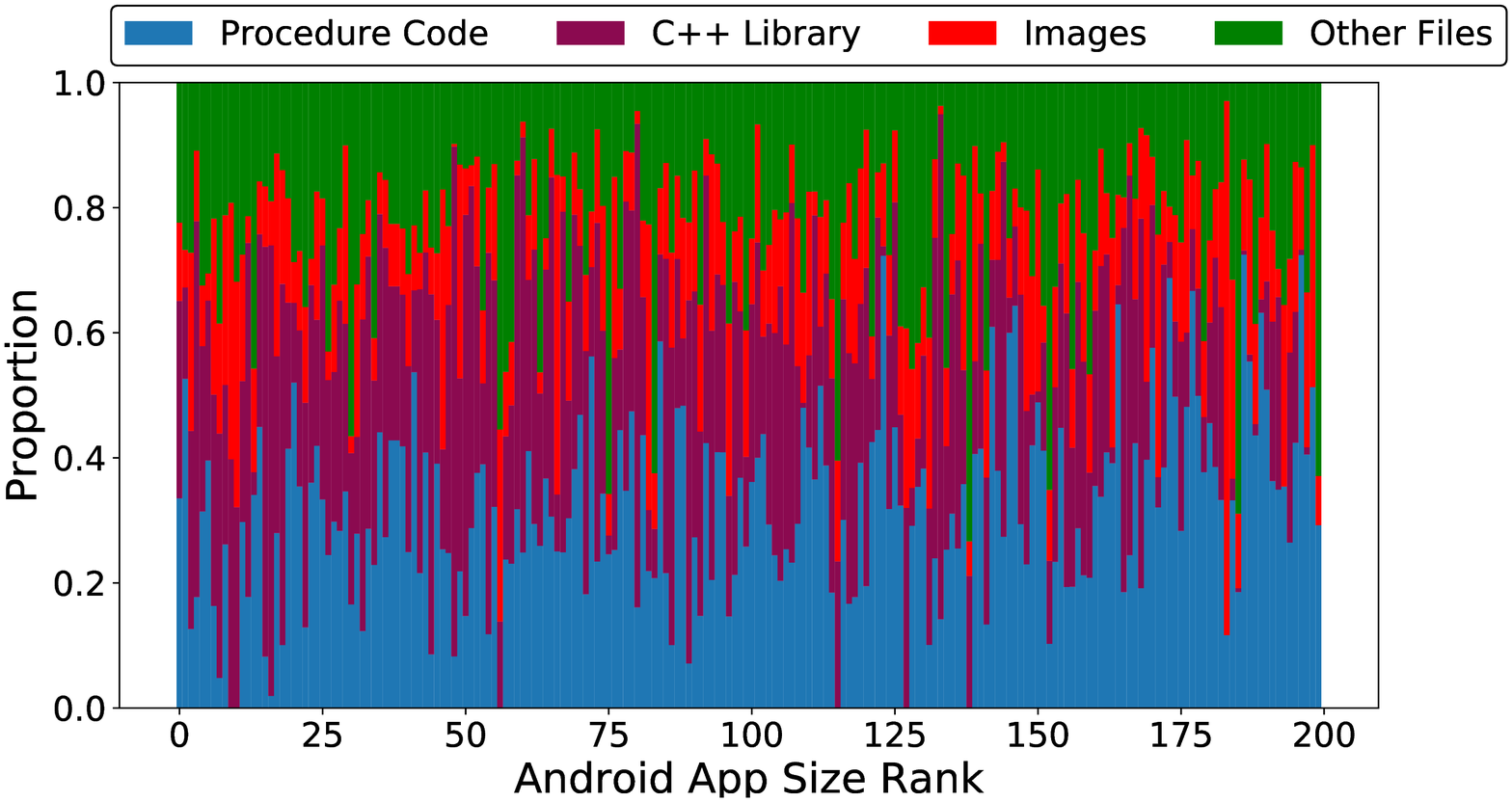}}
	\subfigure[Package composition: Mini-program]{
 		\label{fig:ex1:f} 
 		\includegraphics[width=2.8in]{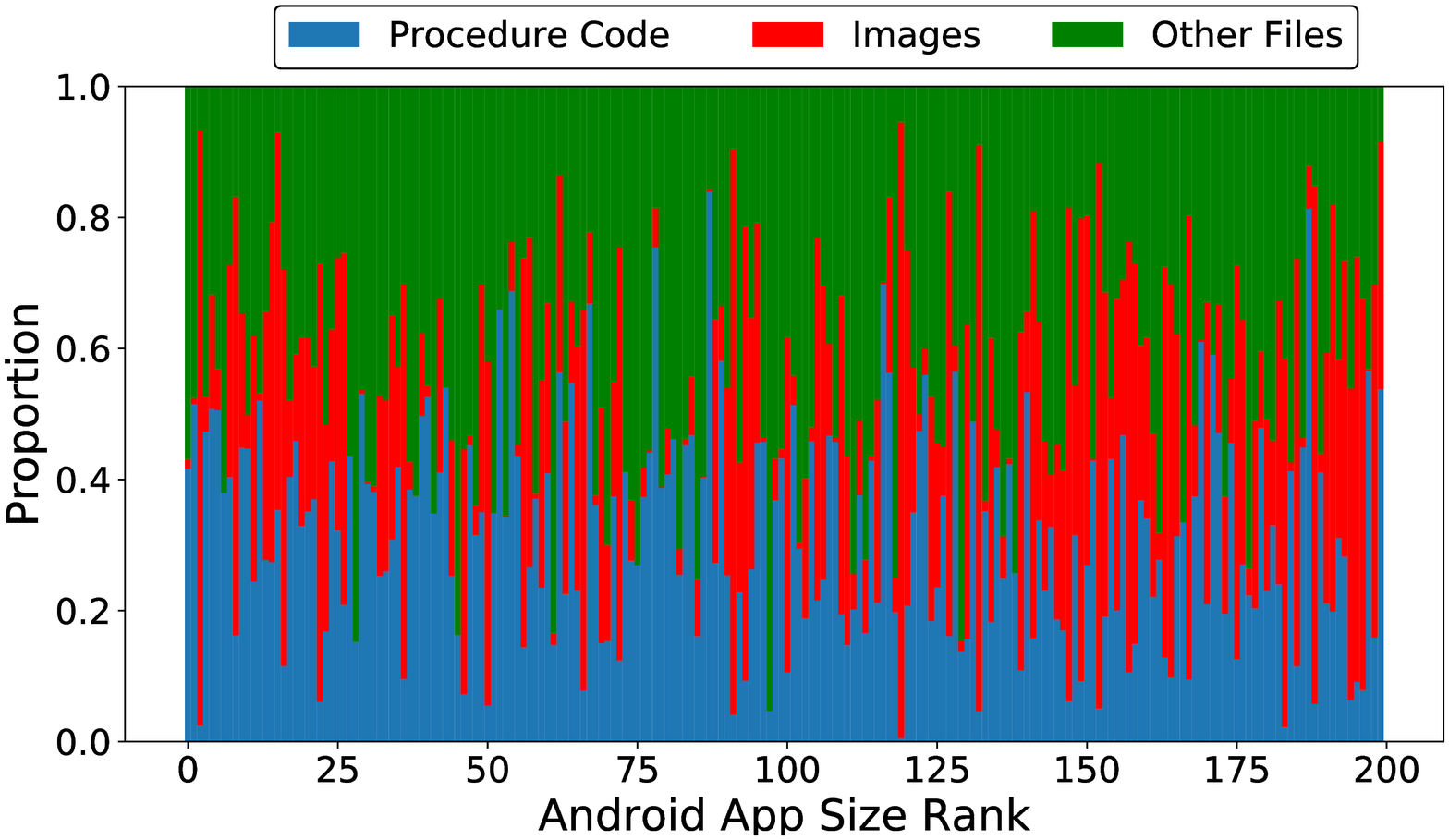}}
	\caption{\small Detailed comparison between 200 pairs of Android app and its mini program version. The pairs are ranked by the descending size of the Android APK package.}
	\label{fig:ex1} 
\end{figure*}
 
For each mini-program and Android app pair, we analyze several key metrics to better understand how the two differ in content, functionality and software package size.

\begin{packed_itemize}\vspace{-0.05in}
\item {\em Installation package size:} the installation size for mini-program (WXAPKG size) and the Android app (APK size).
\item {\em Image size and number of images:} total size of all image files and total number of image files in the respective packages. Applications with richer features tend to have more images.
\item {\em Page count:} a measure of number of features provided by the application.
Since mini-programs register individual pages in their common settings file, {\em i.e.\/ app-config.json}, we use this to count the pages for each mini-program as a measure of features in the program. Android apps register each activity in their \emph{AndroidManifest.xml} file, and we use the number of activities as the measure of features for Android apps. In our experience, mini-program pages correlate roughly with Android activities.

\item{\em Composition of installation package:} the proportion of individual components that make up the installation package, including images, procedure code and support files. 
\end{packed_itemize}\vspace{-0.05in}

We use the reverse engineering tool Apktool~\cite{Apktool} to decompile each target Android app and convert the Dalvik bytecode to Smali\footnote{Smali/Baksmali is an assembler/disassembler for the dex format used by Android's Java VM implementation.} code for our analysis. 


\para{Key Observations.} We plot key results from these metrics in Figure~\ref{fig:ex1}. All graphs are sorted by descending order of the size of the Android app, ranked 1 (largest) to 200 (smallest). Thus all graphs are consistent on the X-axis.
From these results, we make three main observations. 
\begin{packed_itemize}\vspace{-0.05in}
\item Mini-programs and their equivalent Android versions differ significantly in the size of installation package. Mini-programs are 5-50x smaller than their Android counterparts. Surprisingly, there seems to be little or no correlation between package sizes (Figure~\ref{fig:ex1:a}; several large Android apps (tens of MBs in size) translated to some of the smallest mini-programs ($<$100KB).


\item Not surprisingly, Android apps contain much more images, larger images than their mini-program counterparts (Figure~\ref{fig:ex1:b}). Bigger, more feature-rich Android apps lost more features in the translation to their mini-program counterparts. 


\item For most Android apps, program code (procedure code and C++ library) dominates the installation package, often taking 60-70\% of the total footprint. Images only occupy 10-20\%.  Not surprisingly, images often dominate the much smaller installation packages of mini-programs. 
\end{packed_itemize}\vspace{-0.05in}

Considered as a whole, our analysis of popular pairs of mini-program / Android apps shows that current Android apps have used a variety of techniques to generate compact mini-program counterparts, and apps vary widely in how much potential code/resources are available for trimming.
\section{Libraries and Bloat in Mobile Apps}
\label{library_mechanism}
\label{sec:library}

Our results in the previous section identified Java Libraries as a potential culprit for the rapid growth of install packages in Android apps. Here, we take a closer look at how resource files, libraries and code make up the components of an Android app, by examining the code structure of a large range of popular Android apps. 

\para{Android App Dataset.} 
We build a large set of popular Android apps and use it for our code analysis and code-trimming experiments.
We start with a ranking of top free Android apps from the popular app analytics platform App Annie\footnote{https://www.appannie.com/}. We choose top 32 Android app categories from Google Play, and download the top ranked 100 apps in each category, for a total dataset of 3200 apps. Average app size is 22.70 MB.  Popular apps in our dataset include Duolingo, Khan Academy, Walmart, Uber and McDonald's. 

\begin{figure}
	\setlength{\belowcaptionskip}{-0.2in} 
	\setlength{\abovecaptionskip}{0.18in} 
	\centering
	\includegraphics[width=1.8in]{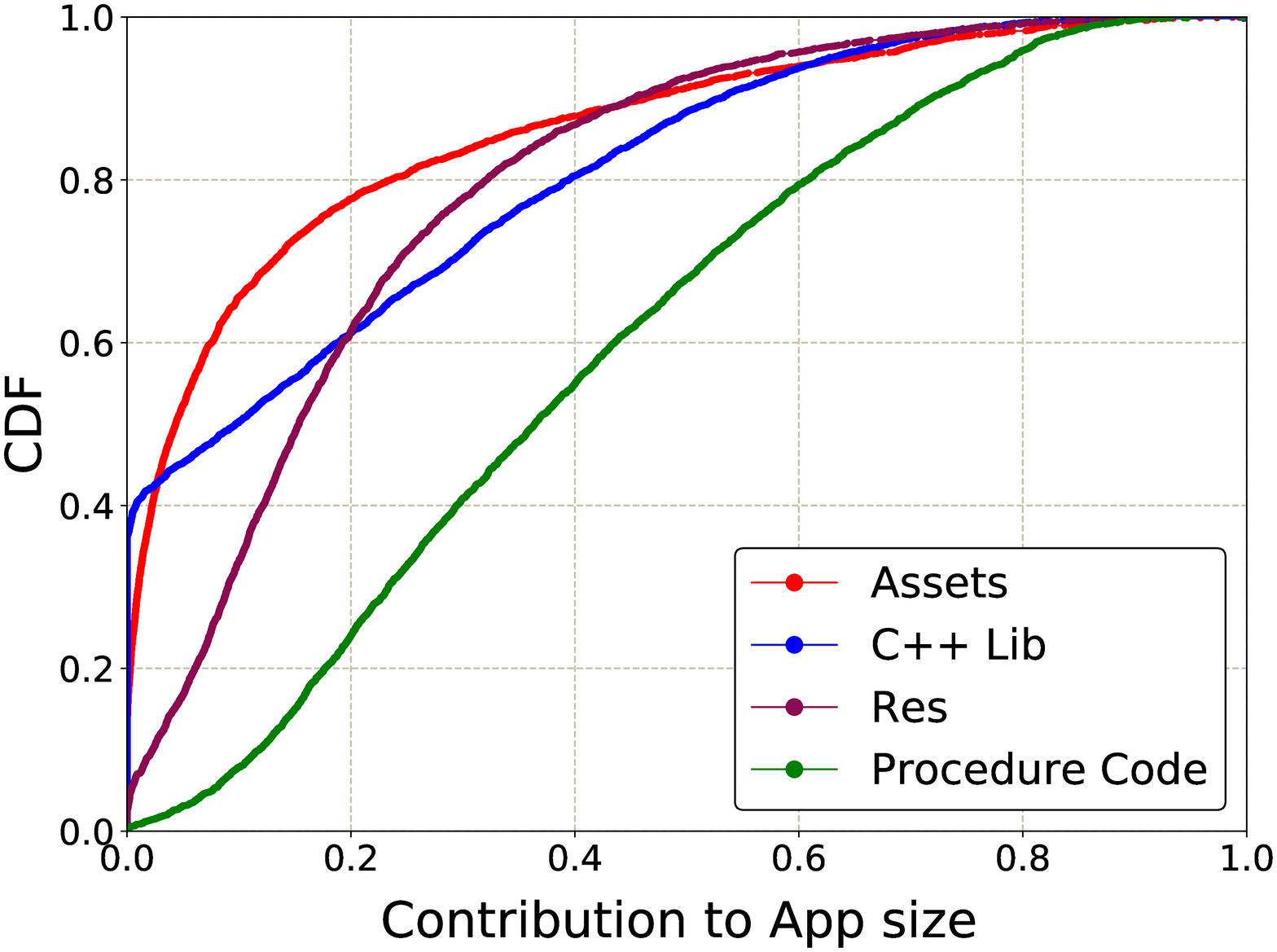}
	 \vspace{-0.1in}
	\caption{\small Contribution of C++ Libraries, Procedure code, Resources and Assets to Android App size.} 
\label{fig:assets}
\end{figure}

\begin{figure*}[t]
	\setlength{\belowcaptionskip}{-0.1in} 
	\setlength{\abovecaptionskip}{-0.02in} 
	\centering
	\subfigure[Absolute code size]{
		\label{fig:lib:a} 
		\includegraphics[width=1.9in]{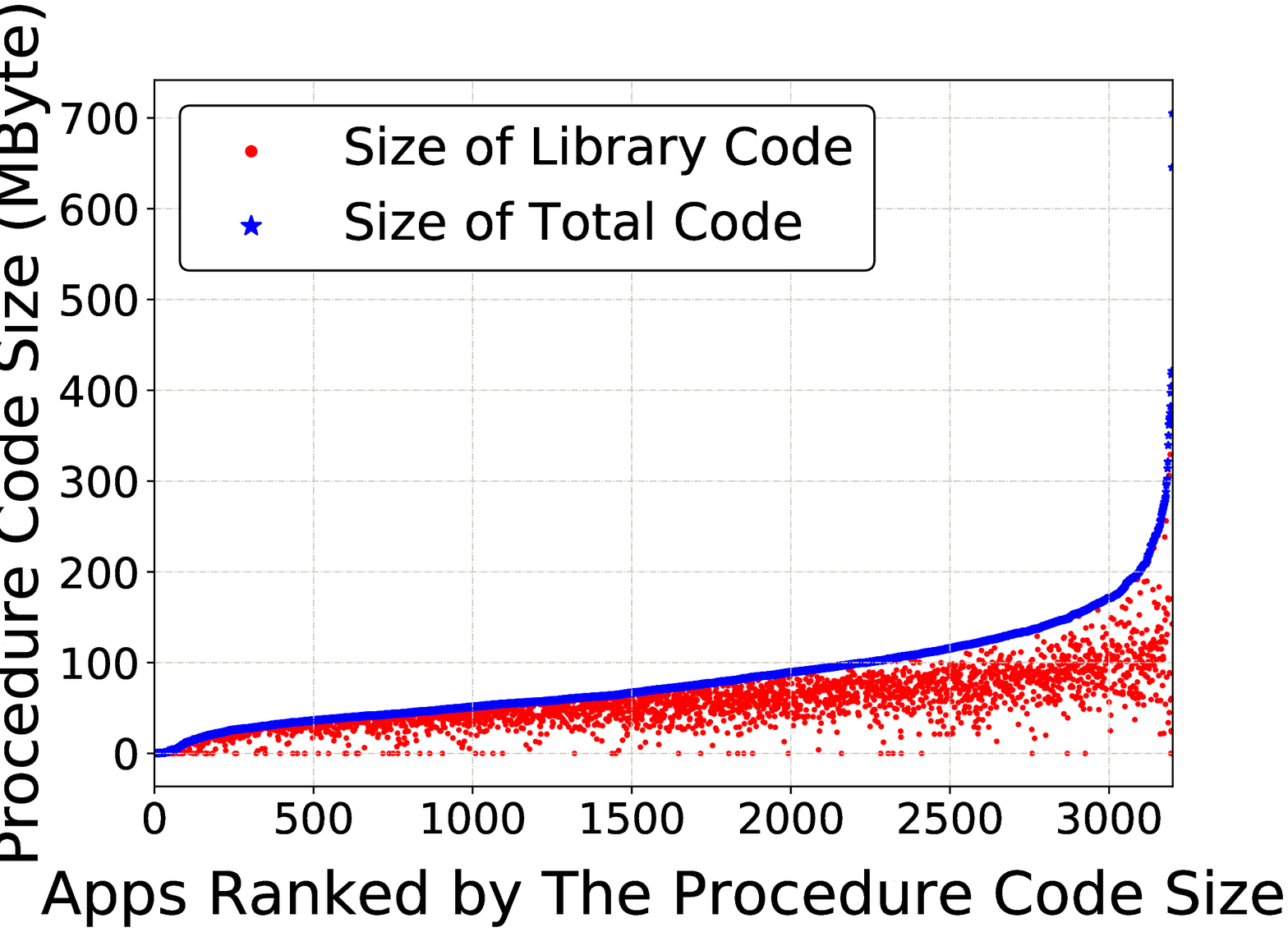}}
   \hspace{0.15in}
	\subfigure[CDF of Java Lib size/Procedure code size ]{
		\label{fig:lib:b} 
		\includegraphics[width=1.9in]{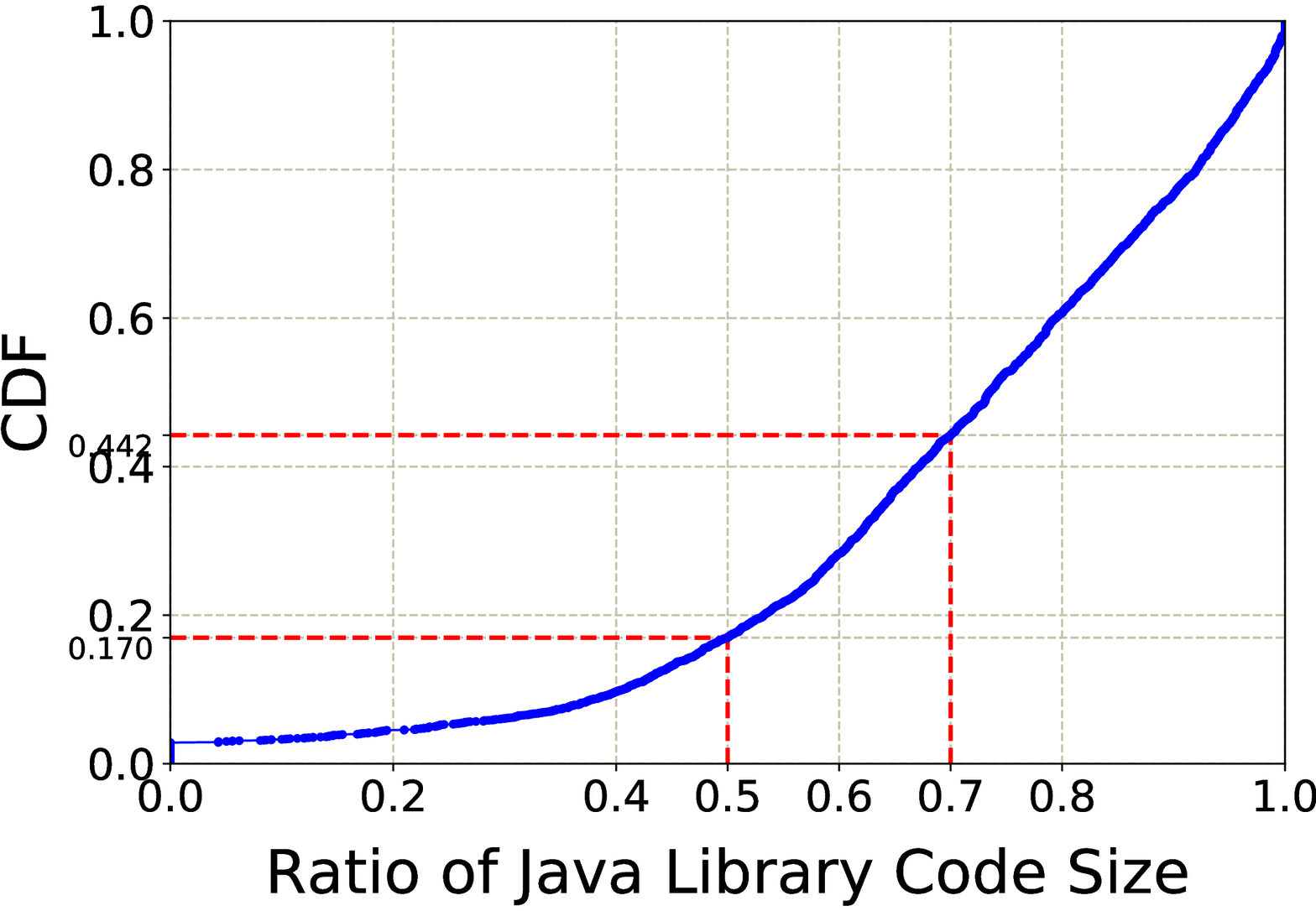}}
	\hspace{0.15in}
	\subfigure[Average Java size ratio across app categories]{
		\label{fig:lib:c} 
		\includegraphics[width=1.9in]{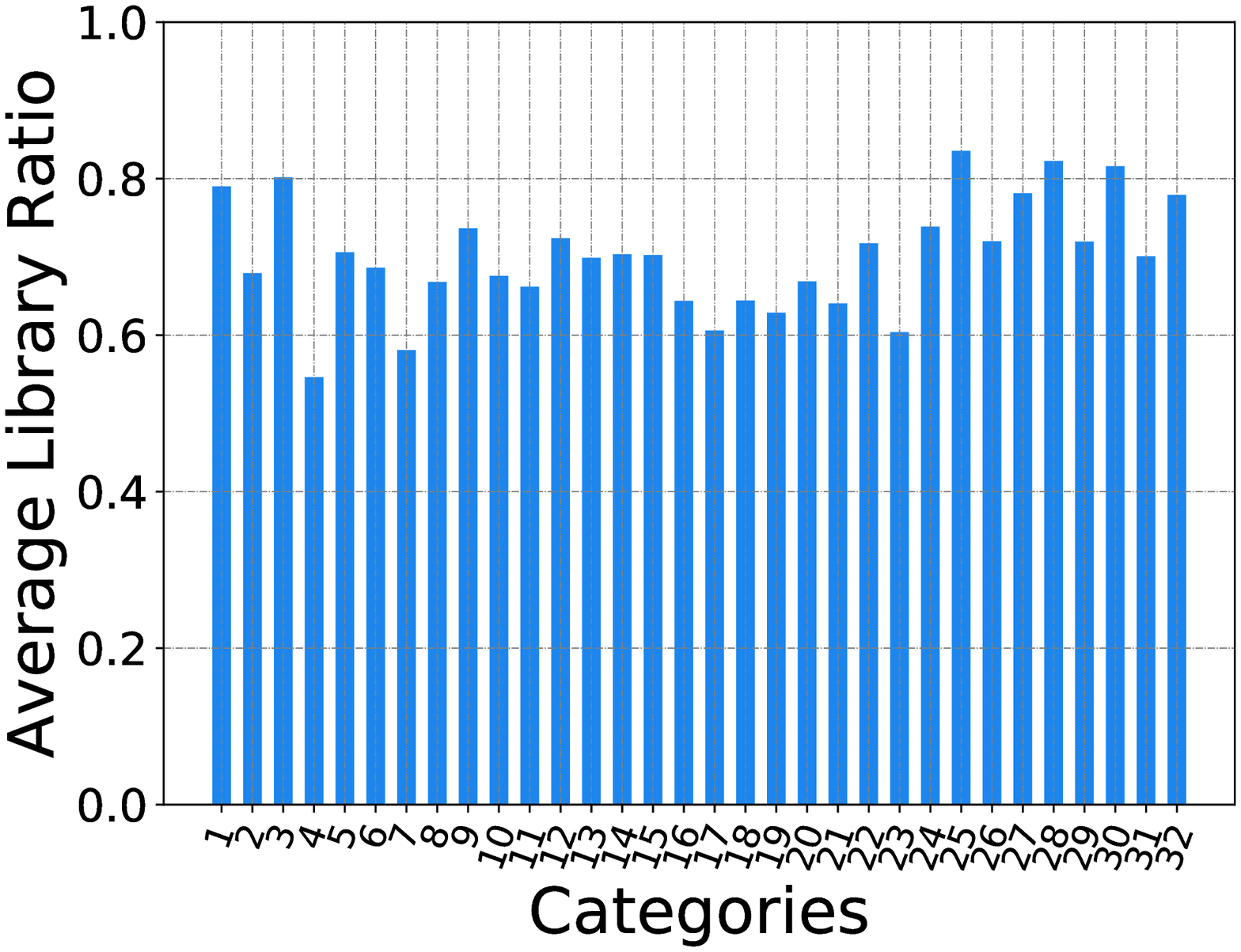}}
		\vspace{-0.15in}
	\caption{\small Comparing the Java library code size to the total procedure code size.	(a) absolute code size of the total procedure code (blue) and that of the Java libraries (red); (b) distribution of (Java library size/Procedure code size); (c) average (Java Library size/Procedure size) across 32 app categories.}. 
	\label{fig:lib} 
\end{figure*}

\begin{figure*}[t]
		\setlength{\abovecaptionskip}{-0.01in} 
	\setlength{\belowcaptionskip}{-0.1in} 
	\centering
		\subfigure[Android App]{
		\label{fig:three:b} 
		\includegraphics[width=2.in]{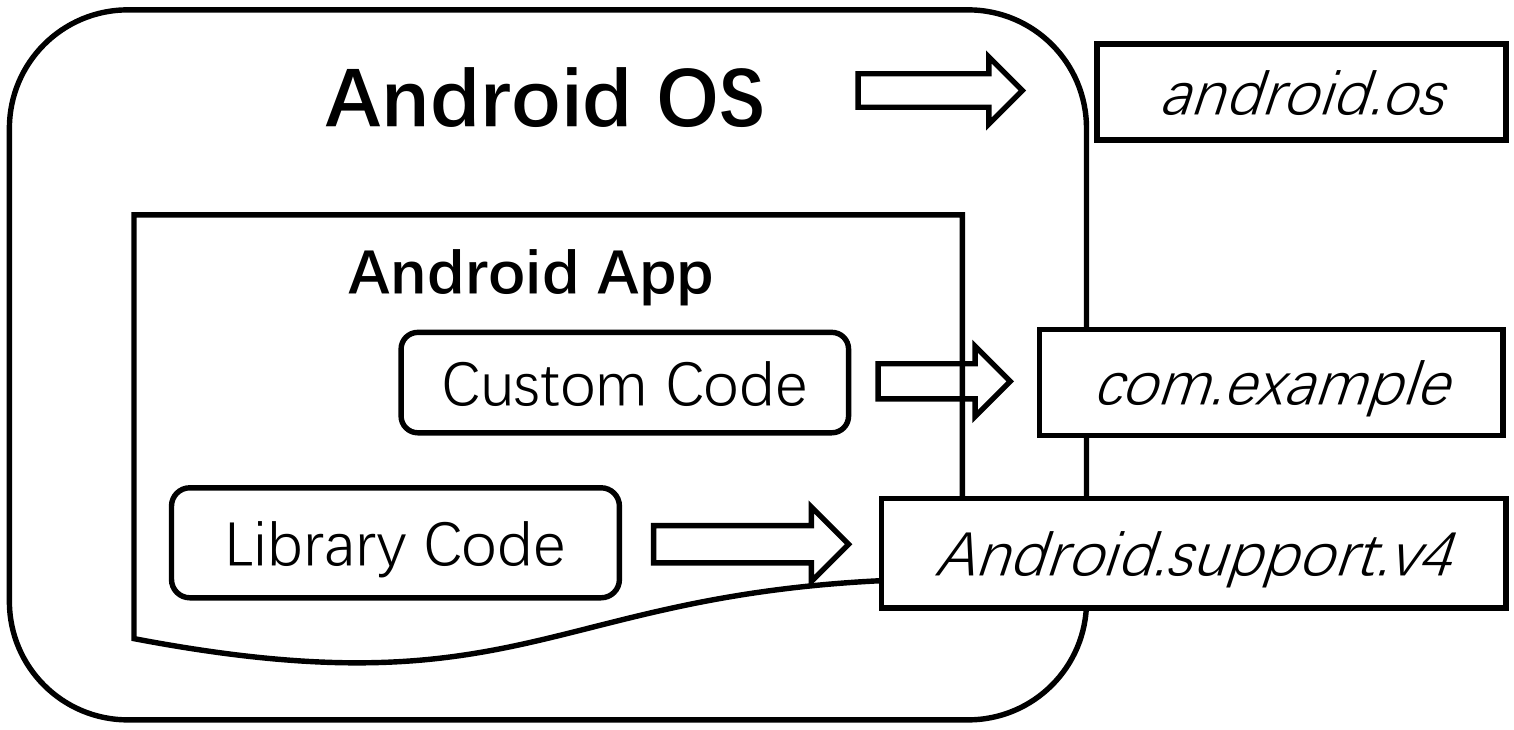}}
			\hspace{0.2in}
	\subfigure[WeChat Mini-Program]{
		\label{fig:three:a} 
		\includegraphics[width=2.in]{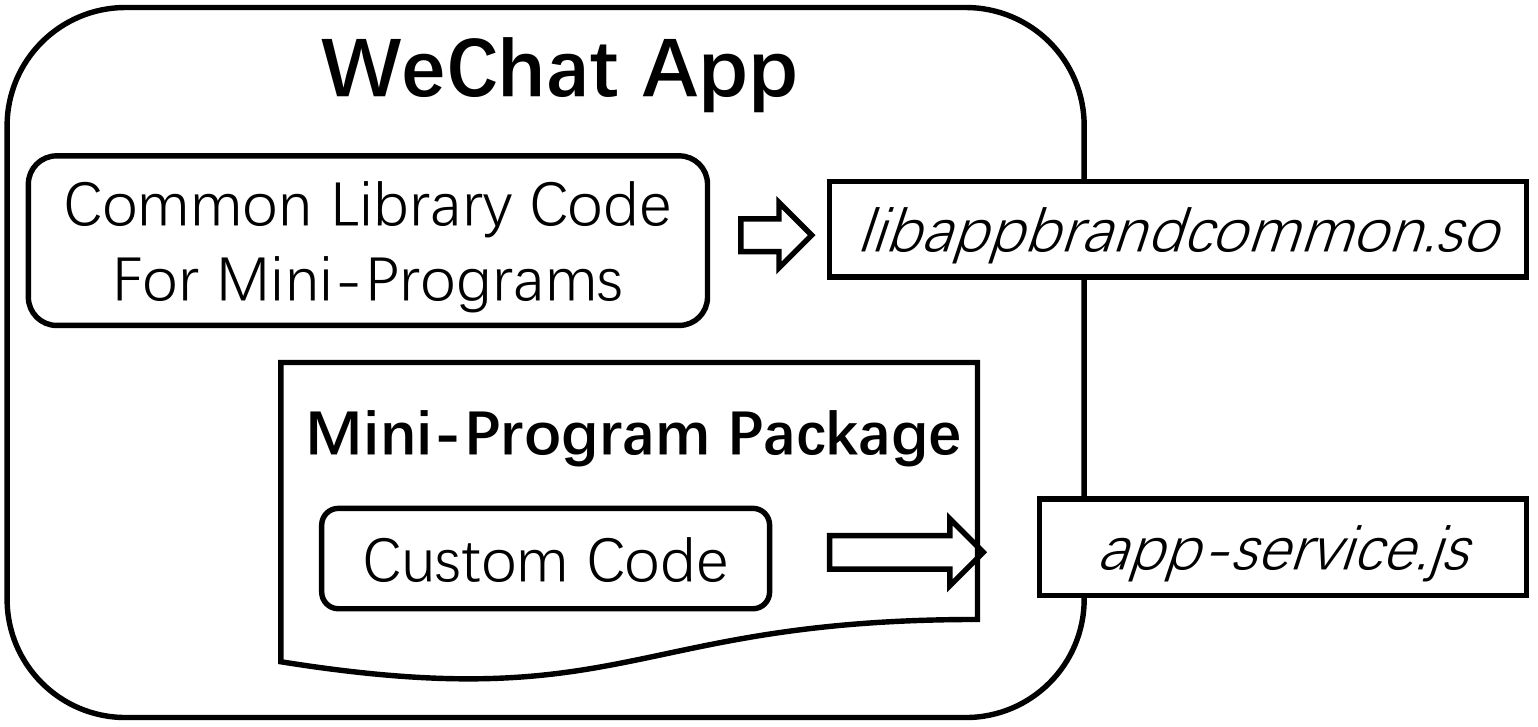}}
		\vspace{-0.1in}
	\caption{\small Library Mechanism of Mini-Program and Android App}
	\label{fig:three} 
\end{figure*}

\subsection{Components of Android Apps}
As we did earlier in Tables~\ref{tab:four} and~\ref{tab:five}, we divide the components of an Android app into four key categories:  {\em Resource files}, {\em Assets}, {\em C++ Library files}, and {\em Procedure code files (including Java Libraries)}. Resource files (files in the {\em{res}} directory) and Assets (files in the {\em{assets}} directory) both include images and supplementary content, but differ in how they are used by the app. Assets tend to cover documentation, icons, and other multimedia files, while resource files are accessed by the app in memory via resourceIDs. C++ Library files are external libraries accessed via Java Native Interface (JNI), and procedure code files represent both core Java code and Java Libraries. 



Figure~\ref{fig:assets} shows how these four components contribute to the app size, across the 3200 Android apps in our dataset. The exact contribution per component varies across Apps, but the procedure code file is generally the biggest contributor. Fortunately, C++ Library files (which are likely to be the hardest to modify or trim) are reasonably small contributors to code size on most apps. 
Instead, it seems there is ample opportunity to optimize procedure code files, along with resources in {\em{/res}} and resources in {\em{/assets}}.

\subsection{Impact of Java Libraries}
We observed earlier that Java libraries can add substantially to the code size of an Android app. Here, we 
study how Java library code (a sub-component of procedure code) contributes to overall package size in Android apps. Java libraries can be further classified as official or third-party libraries. Many Android apps use third-party libraries, {\em e.g.\/}, advertising service libraries to generate revenue~\cite{ad_detect_mobisys15}.  The official libraries are offered by Google, and can be identified by their names, {\em e.g.\/}, Android.support.v4. We detect third-party libraries using an existing framework called LibRadar~\cite{Ma_ICSE16}.

Figure~\ref{fig:lib} quantifies the code size of Java libraries from different perspectives. Figure~\ref{fig:lib:a} plots, for each of our 3200 apps sorted by procedure code size, the absolute size of total procedure code (blue dot) and the absolute size of Java Libraries (red dot). For both, the code size is the size of the Smali files obtained after decompilation. Our key observation here is that for the overwhelming majority (96.7\%) of apps, procedure code is dominated by Java libraries. This is further confirmed in Figure~\ref{fig:lib:b}, which plots the CDF of the ratio between Java library code size and total procedure code size. For more than 55\% of apps, Java library code accounts for more than 70\% of total code size. Non-Java library code makes up the majority of procedural code in only 17\% of apps. 
Figure~\ref{fig:lib:c} shows that this dominance by Java libraries is consistent across app categories. 



\subsection{Library Management in Android vs. Mini-Programs}
While procedure code in Android apps is dominated by Java libraries, code in mini-programs are not dominated by their libraries. This can be directly attributed to how libraries are managed by WeChat mini-programs and their Android counterparts. Figure~\ref{fig:three} illustrates the two library management mechanisms, which we describe next. 

\para{Android apps.} Each Android APK includes both library code and app-specific code (codes written by app developers). For example, \emph{Android.support.v4} is a library package and \emph{com.example} is a custom code package. When generating an APK, all library codes and custom codes are packed into the same APK. 


\para{WeChat mini-programs.} 
The library file used by WeChat mini-programs is the \emph{libappbrandcommon.so} file in the lib dictionary of the WeChat app.  When generating a mini-program, only custom codes are packed into the mini-program, and not library files. In other words, mini-programs do not include  library files in its installation package, but uses the library file included in the WeChat app. 

This key difference likely arises from the lack of consistency in application libraries across Android devices. Apps packing their own library code improves robustness and increases the likelihood of the app running on different devices and Android versions. The price for this robustness is redundant library code packed into the APK files of each Android app. In contrast, the consistency offered by WeChat's own mini-program platform means mini-programs can make stronger assumptions about library versioning, and a single common library can be shared across all mini-programs. This dramatically reduces the duplication of library code across apps or mini-programs.

\section{Trimming Code and Resources on Android Apps}
\label{sec:trim}

Our early analysis of the Weather mini-program in Section~\ref{section3} showed that some apps included large (and likely unused) Java libraries in their install package.  The size of these Java libraries could account for significant code size discrepancy between Android apps and mini-programs. Our additional hypothesis is that most Android apps only use a small subset of modules in Java libraries, but developers often import the entire library because manually identifying the code snippets for the target modules is labor-intensive. As a result, significant portions of the package code is actually unused by the app, i.e. ``code bloat,'' and should be trimmed.

In this section, we propose a systematic framework to trim Android apps: including identifying code bloat, removing it, and then repacking the app. We can perform similar operations to identify resource bloat ({\em e.g.\/} unused images) and remove them from the app. 
Here, we describe our proposed process for trimming program code (\S\ref{subsec:trimcode}) and resources (\S\ref{subsec:trimresources}), and a process that integrates both. Later in \S\ref{sec:eval}, we evaluate the effectiveness of our proposed app trimming techniques and usability of the trimmed apps. 




\subsection{Trimming Code Bloat}
\label{subsec:trimcode}
Our trimming framework consists of four sequential steps:  preprocessing, code decompilation, code bloat detection, and app repacking and validation. It takes as input an Android installation package, and outputs a repacked app with code bloat removed. More specifically, we first preprocess the input app, then unpack it using the dex2jar tool where Dalvik bytecode gets transformed to Java bytecode. We then leverage Proguard~\cite{porguard}, a Java class file shrinker, optimizer and obfuscating tool to identify and trim code bloat. Finally, we repack the app and validate that its functionality has not been disrupted by the trimming process.  We now describe these steps in detail. 



\para{Preprocessing.} The goal of preprocessing is to identify Android apps that cannot be decompiled and repacked due to built-in security mechanisms (e.g. encryption or code signatures) that prevent code modification or decompilation~\cite{ad_detect_mobisys15} (more discussion in \S\ref{discussion}). The preprocessing step tests Android apps by first re-signing the app (as a different developer from the original) and check if it can still run properly, and if successful, then decompiles the app using bytecode transformation, repacks the app, and then re-signs the app. If the re-signed app passes both tests, it is suitable for code trimming. Note that this limitation only applies because we are an untrusted third party. Google or an authorized third party could use authenticated tools to bypass an app's protection mechanisms and enable code trimming.


\para{Identifying and Removing Code Bloat.} 
To identify code bloat, we first apply the dex2jar tool to convert the app's Dalvik bytecode to Java bytecode.  Here the conversion supports both apps with single-dex and multi-dex.  For apps with multi-dex, we merge the Java bytecodes per DEX file by their file paths and use a map file to record the file path that will later be used by the re-pack step.  

Next we use the ProGuard tool to explore the app execution space, recursively searching for any class and class members that the app will actually use. Those not found by ProGuard are treated as code bloat and removed from the app package.  The search of the app execution space requires a seed or entry point. For this we use {\em MainActivity}, the actual entry point of the target Android app, accessible directly from the global configuration file ({\em AndroidManifest.xml}).  To be conservative, we also do not trim any subclass of the Application, Activity, Service, BroadcastReceiver and ContentProvider class, and instead use them as extra entry points. Furthermore, because ProGuard only targets Java, we do not trim the Enum class, Java Native Interface and construction methods as well as widgets used by the xml files.  Finally, our current search implementation does not consider Java reflection and dynamically loaded code instantiated by the Java class loader, because ProGuard does not recognize them~\cite{Sif2013}. This means we could accidentally trim useful code, but we can identify any such mistakes and recover during the app validation step.

\lstset{style=mystyle, escapeinside=``, aboveskip=1em}
\begin{lstlisting}[language=Java,style=mystyle,caption={\small The original MainActivity},label=code:init]
		public class MainActivity extends AppCompatActivity {
			protected void onCreate(Bundle savedInstanceState) {
				super.onCreate(savedInstanceState);
				setContentView(R.layout.activity_main);
				int a = 1, b = 2, c = 3, d; d = sum(a, b); }
				int sum(int num1, int num2){ return num1 + num2; }
				int sub(int num1, int num2){ return num1 - num2; }
		}
\end{lstlisting}

\lstset{style=mystyle, escapeinside=``, aboveskip=1em}
\begin{lstlisting}[language=Java,style=mystyle,caption={\small MainActivity bytecode after decompilation},label=code:decompilation]
public class MainActivity extends AppCompatActivity {
	public MainActivity() {}
	protected void onCreate(Bundle var1) {
		super.onCreate(var1);
		this.setContentView(2131296283);
		int var2 = this.sum(1, 2);         }
	int sum(int var1, int var2){ return var1 + var2;}
	int sub(int num1, int num2){ return num1 - num2; }
}\end{lstlisting}
\lstset{style=mystyle, escapeinside=``, aboveskip=1em}
\begin{lstlisting}[language=Java,style=mystyle,caption={\small MainActivity bytecode after code trim},label=code:shrink]
public class MainActivity extends AppCompatActivity {
	public MainActivity() {}
	protected void onCreate(Bundle var1) {
		super.onCreate(var1);
		this.setContentView(2131296283);
		int var2 = this.sum(1, 2);         }
	int sum(int var1, int var2) {return var1 + var2;}
}
\end{lstlisting}

{\em An Illustrative Example.} Here is an example on how to identify and remove code bloat from an app. Listing~\ref{code:init} shows the Java code of the MainActivity class in a sample Android app, where ``onCreate'' sets the layout file of the main page.  Since onCreate is the entrance to the program, the \emph{sub} function will not be used after the program is executed, and should be trimmed. 
Listing~\ref{code:decompilation} shows the MainActivity Java bytecode after decompilation, where variable names are replaced by their values, and any unused variables ({\em i.e.\/} $c$) are removed. Listing~\ref{code:shrink} shows the result after code trim, where the unused function \emph{sub} is removed.


\para{Repacking and App Validation.} After re-packing the trimmed app, we need to validate if it still functions correctly.  For this we follow existing works~\cite{ad_detect_mobisys15,Privacy_Smartphones_Ubicomp17} and run for three minutes an automatic UI traversal script based on the Appium~\cite{appium} and Monkey~\cite{monkey} scripts. This script will validate the functionality of the trimmed app.



It is worth noting that another option for app validation is the PUMA tool~\cite{PUMA_Mobisys14}. Unfortunately, PUMA only supports up to Android 4.3, and a significant portion of apps (roughly one third of apps tested) fail to install on the Android 4.3 emulator (due to SDK limitations). This makes PUMA unsuitable for our final app validation. But for the 538 apps in our dataset that pass the preprocessing test and can run on Android 4.3, validation using PUMA shows that 486 apps (90.33\%) function properly after code trimming, which is consistent with our results in \S\ref{sec:eval}.

\subsection{Trimming Resource Bloat}
\label{subsec:trimresources}
We also seek to detect and remove unnecessary bloat in resource files, {\em i.e.\/}, both Res resources and Assets. For this we use static code analysis to identify unused resources in the app, from images to XML files. Specifically, we first use Apktool to decompile the target Android app for static code analysis, which converts the Dalvik bytecode to Smali code and parses the resource file. 
Parsing the resource files allows us to identify unused resource files.  





\para{Identifying Bloat in Res Resources.}  The {\em res} directory contains different file types like drawable, string, color, etc. We only identify bloat in drawable resources like images and XML files, because trimming other resource types requires modifying the XML file, and can potentially disrupt the decompilation and repacking process. 


First, parsing resource files will produce a (\emph{public.xml}) file in the folder \emph{res/values}, which records every res resource's ID, name and type (drawable, string, attr, color, array, etc). As we mentioned in \S\ref{library_mechanism}, after they are compiled, any res resource is accessed through its resource ID. To identify all resources used by the app, we can just search for them in each Smali file\footnote{Here we need to exclude any res resources found in the R class Smali files, since those files include all res resources.}.


\begin{figure*}[t]
	\setlength{\belowcaptionskip}{-0.1in} 
	\centering
	\includegraphics[width=4.5in]{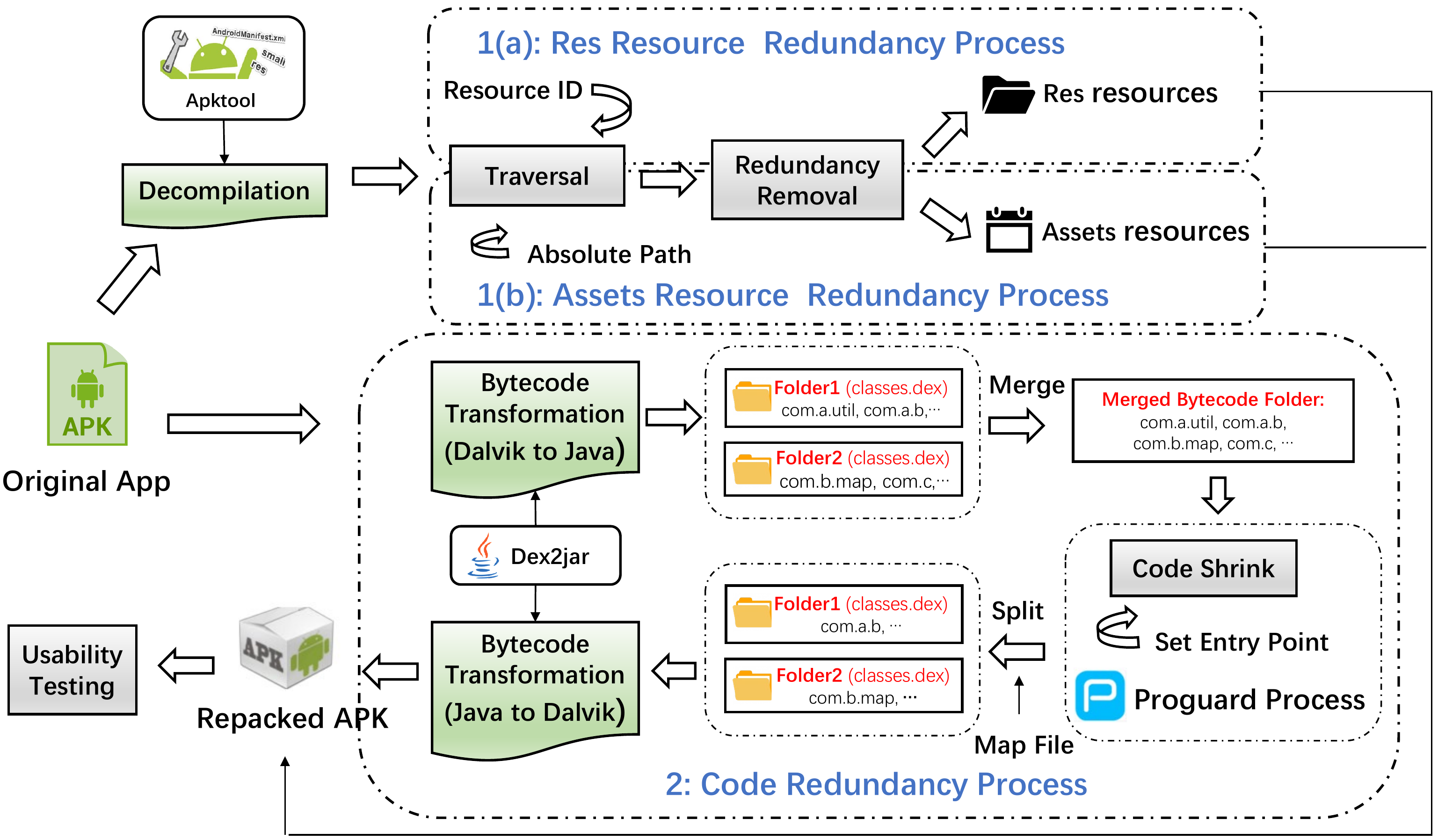}
	\caption{The overall process of our proposed app trimming framework.}
	\label{fig:five} 
\end{figure*}

\para{Identifying Bloat in Assets.} Assets usually store static resources like database files and videos, which are neither code nor configuration files.  Thus resources in assets are not compiled when packed into an APK.  Since asset resources are accessible by their absolute path in the code, we can identify them by traversal searching the absolute path of each asset resource in each Smali file. Resources not identified by this search are trimmed. 


\subsection{Putting Everything Together}
Finally, we can integrate the code trimming process with the resource trimming process to build a fully automated app trimming framework for Android apps. Our code is available on GitHub (link removed to preserve anonymity). The overall framework is shown in Figure~\ref{fig:five}. It trims the assets, the res resources, and finally the procedure code in sequence. 




\section{Evaluation} 
\label{sec:eval}
In this section, we evaluate the performance of our proposed app trimming framework. We consider two key metrics: effectiveness as measured by  reduction in mobile app size, and correctness in terms of whether the trimmed app still functions properly. 

\para{Experimental Configuration.} Our evaluation considers the 3200 top Android apps described in \S\ref{library_mechanism}. To experiment with this wide range of apps~\cite{android8_mobisys18}, we install these 3200 apps on an Android emulator (Samsung Galaxy S7, Android 8.0), and ran the emulator on two identical Ubuntu 16.04 machines with 6-core 3.60GHz CPU and 100GB memory.  76 out of the 3200 apps fail to install on the emulator, while 204 apps fail to run properly after installation. We removed them from our experiments. In the end, our experiments used the remaining 2920 apps to test the effectiveness and correctness of our trimming framework.

\begin{figure*}[t]
	\setlength{\abovecaptionskip}{0in} 
	\setlength{\belowcaptionskip}{-0.15in} 
	\centering
	\subfigure[Absolute app size reduction]{
		\label{fig:final:a} 
		\centering
		\includegraphics[width=2in]{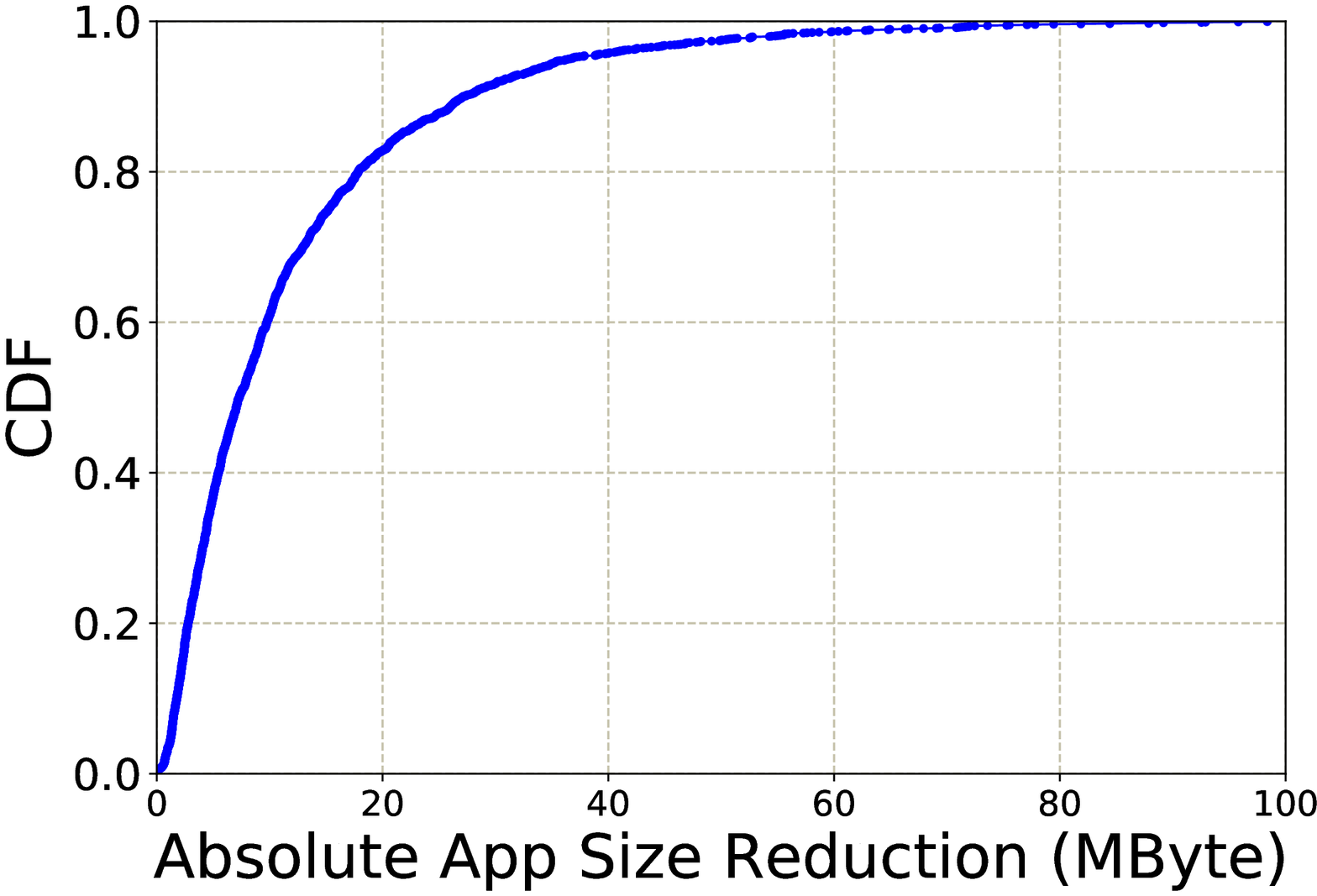}}
	\subfigure[Normalized app size reduction]{
		\centering
		\label{fig:final:b} 
		\includegraphics[width=2in]{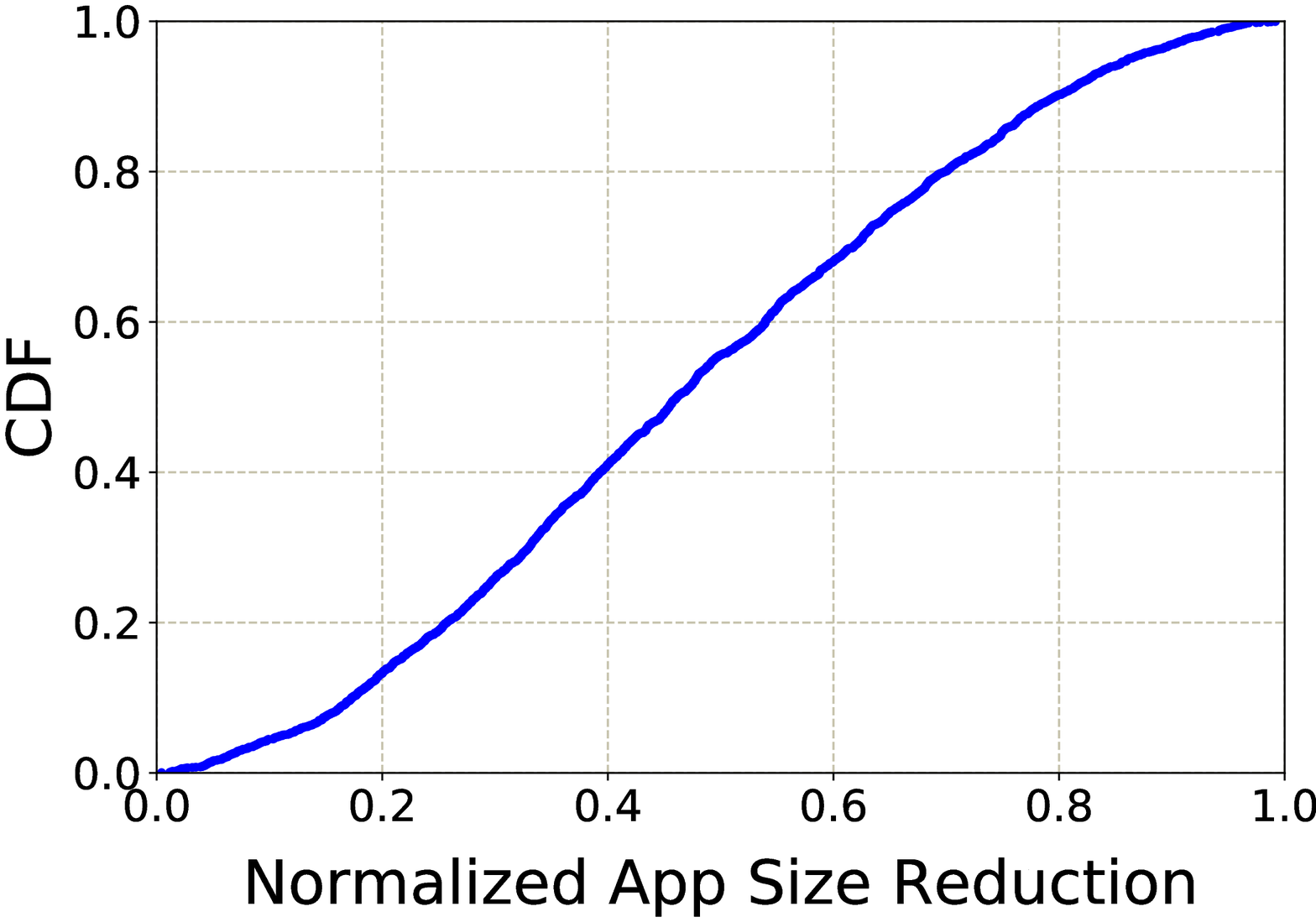}}
	\subfigure[Normalized amount of components trimmed]{
		\centering
		\label{fig:final:b} 
		\includegraphics[width=2in]{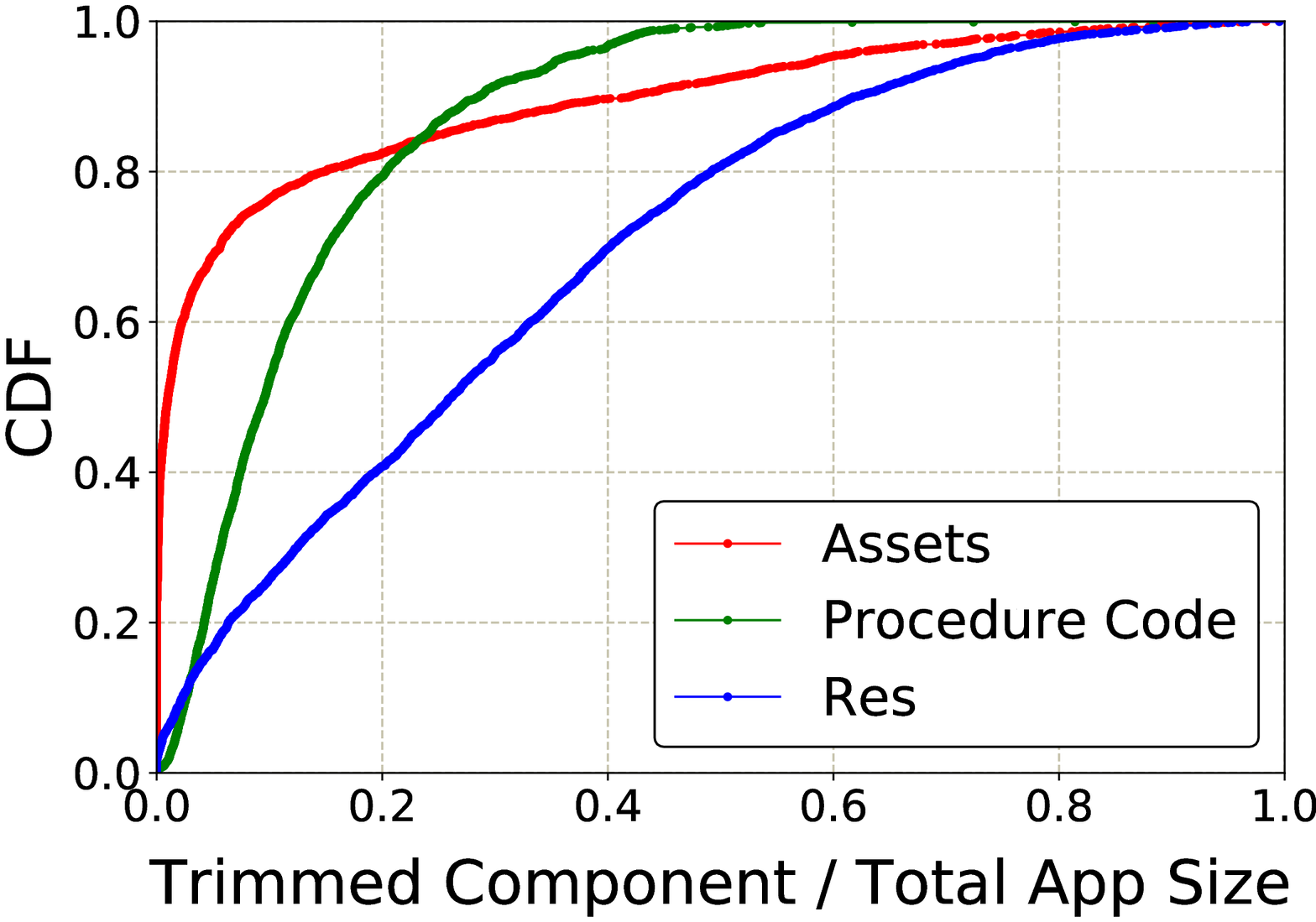}}
	\caption{\small Our automated, app trimming framework can effectively reduce app size.}
	\label{fig:final} 
\end{figure*}

\subsection{Effectiveness of App Trimming} 
Figure~\ref{fig:final} plots the CDF of the absolute app size reduction, the normalized app size reduction and the per component reduction normalized by the app size. From Figure~\ref{fig:final}(a-b), we see that for 40\% of the apps, trimming the app can reduce the app size by at least 10MB, or at least 52\%.  Here are some specific examples: Duolingo reduces from 19.87MB to 12.07MB, Khan Academy reduces from 21.94MB down to 16.48MB, Uber reduces from 60.64MB to 31MB, and McDonald's reduces from 42.5MB to 15MB.  Figure~\ref{fig:final}c further shows that trimming res resources (images) is highly effective, followed by trimming procedure code (Java library files).



These results confirm that our design can effectively and significantly reduce the size of Android apps by trimming code and resource bloat.  
Our trimming process is fully automated, allowing third-parties to easily generate lightweight mobile apps for developing regions without sacrificing basic functionality. For app developers, our framework helps to identify potential code and resource bloat for performance optimization. 


\subsection{Correctness of App Trimming}
Of the 3200 Android apps we tested, 2920 apps passed our preprocessing steps and were deemed suitable for automated trimming. Of these, 2617 apps (89.62\%) passed validation and operated correctly after trimming. Most apps observed significant drops in their Java code sizes, with reductions ranging from 60-80\% of their original size. More than 70\% of apps saw a drop in redundant code of more than 5MB, and a small number of apps saw a size reduction of more than 20MBs after automated trimming.

\section{Discussion}
\label{discussion}

\para{Reducing redundancy during app development.} Results of our study show that significant code and resource redundancy are widely present in today's mobile apps. Removing or limiting them during app development is quite feasible using existing tools.  

To reduce code redundancy, one potential tool is Proguard~\cite{porguard}, which has already been integrated into Android Studio~\cite{android_studio}, the official development environment for Android. Developers can easily edit Proguard configurations to remove redundant code in their project. However, it is also possible that developers who want to prevent code trimming could use tools like Proguard to intentionally obfuscate their code. For example, developers updating their apps over time might opt to save code belonging to deprecated features rather than removing them fully, since code removal might introduce bugs that require more effort to locate and fix.


Tools also exist for removing resource redundancy during app development.  Android Lint is a a code scanning tool provided by Android SDK and integrated into Android Studio. It helps developers identify and correct issues like unused resources during development. Our observations of high levels of resource redundancy likely indicates that few developers are using Lint. In roughly half of Android apps, more than 50\% of asset resources are redundant, and it is even worse for res resources: in roughly half of Android apps, more than 80\% of res resources are redundant. Since removing unused resources is less likely to produce complex failure modes, developers looking to trim bloat should start with resources.

\para{Lightweight app platforms (for developing regions).} We show that WeChat mini programs exhibit significant size savings when compared to Android apps.  Part of this comes from  mini-programs' shared library access, as discussed in \S\ref{library_mechanism}. This suggests that a platform for lightweight (mobile) applications might benefit mobile users in developing regions, using WeChat's platform as a reference.  Placing commonly used code into a shared, consistent app library would greatly reduce code redundancy  for mobile apps. 

\para{Limitations.} Finally, we discuss limitations of our study.  First, app decompilation and repacking is known to be fragile~\cite{ad_detect_mobisys15}. Errors can creep into the system during the use of reverse-engineering tools like dex2jar, Apktool and enjarify~\cite{enjarify_tool}. For instance, a \emph{unknown opcode} exception was reported when we used dex2jar to translate Dalvik bytecode to Java bytecode. This is because that most reverse-engineering tools read bytecode linearly and the parse process fails when encountering an invalid bytecode.  Thus developers can prevent third-party code trimming by intentionally or accidentally inserting invalid bytecodes into the Android Dex file. Similarly, steps like parsing procedure codes can be disrupted or slowed using unexpected inputs in procedure code. Finally, developers can always use encryption or code signatures to prevent or detect alterations to their code.

\section{Conclusion and future work}
In this paper, we take  an empirical approach to analyzing sources of bloat in today's mobile applications.  Using Wechat mini-programs as a basis for comparison, we were able to identify a number of potential causes for the rapid growth in mobile app package sizes. This in turn, allowed us to identify techniques to significantly reduce size of existing Android applications by modifying and trimming unnecessary code and resources.  

While our techniques have demonstrated significant success in our tests, we believe they represent only initial steps by which developers can support mobile users in developing regions. For example, our work helps to address the challenge of downloading and updating apps in bandwidth-constrained networks. But many mobile apps today make strong assumptions about the availability of network bandwidth, and either fail to operate fully under constrained conditions, or aggressively consume bandwidth to the detriment (and high costs) of their users. We hope our work and others will lead to treatment of bandwidth-constrained networks as a first class consideration, along with development of tools and platforms that more easily integrate support for low-bandwidth networks into a wide-range of mobile applications.

\begin{small}
\bibliographystyle{acm}
\bibliography{trim}
\end{small}

\end{document}